\documentclass[10pt,reprint,twocolumn,aps,showpacs,a4paper]{revtex4-1}

\usepackage{amsmath}
\usepackage{amssymb}
\usepackage{amsfonts, epsfig}
\usepackage{graphicx}
\usepackage{setspace}
\usepackage{calc}
\usepackage{floatflt}
\usepackage{overpic}

\usepackage{color}

\def\e{\varepsilon}

\def\ket#1{|#1\rangle }
\def\bra#1{\langle#1 | }

\def\expect#1{\langle#1 \rangle}

\def\non{\nonumber \\ }
\def\w{\omega}
\def\k{\vec{k}}
\def\r{\vec{r}}
\def\punkt{\;\; .}
\def\komma{\;\; ,}
\def\Tr#1{\textrm{Tr}\left[#1\right]}
\def\ul#1{\underline{#1}}
\def\mat#1{\ul{\ul{#1}}}

\def\kr{k_\mathrm{F} R}
\def\Simp{\vec{S}_\mathrm{imp}}
\begin{document}

\title{Spatial and temporal propagation of Kondo correlations}

\author{Benedikt Lechtenberg}
\author{Frithjof B. Anders}
\affiliation{Lehrstuhl f\"ur Theoretische Physik II, Technische Universit\"at Dortmund, 44221 Dortmund,Germany}

\date{\today}

\begin{abstract}
We address the fundamental question how the spatial  Kondo correlations are building up in time
assuming an initially decoupled  impurity spin $\vec{S}_{\rm imp}$. We  investigate
the time-dependent spin-correlation function $\chi(\vec{r},t) = \expect{\vec{S}_{\rm imp} \vec{s}(\vec{r})}(t)$
in the Kondo model with antiferromagnetic and ferromagnetic
couplings where  $ \vec{s}(\vec{r})$ denotes the spin density  of the conduction electrons  after switching on the Kondo
coupling at time $t=0$. We present data obtained 
from a time-dependent numerical renormalisation group (TD-NRG) calculation.
We gauge the accuracy of our two-band NRG by the spatial sum-rules of the equilibrium correlation
functions and the reproduction of the analytically exactly known
spin-correlation function of the decoupled Fermi sea.
We find a remarkable building up of Kondo-correlation outside of the light cone
defined by the Fermi velocity of the host metal. By employing a
perturbative approach exact in second-order of the Kondo coupling, we 
connect these surprising correlations to the intrinsic spin-density entanglement 
of the Fermi sea. 
The thermal wave length  supplies a cutoff scale at finite temperatures beyond which correlations are exponentially suppressed. We present data for the frequency dependent
retarded spin-spin susceptibility and use the results to calculate the real-time response of a weak 
perturbation in linear response: within the spatial resolution no response outside of the light cone is found.
\end{abstract}



\maketitle

\section{Introduction}

A localized spin interacting antiferromagnetically with a metallic
host is one of the fundamental problems in theoretical condensed mater
physics.  Originally proposed by
Kondo \cite{Kondo1964,*Kondo2005} for understanding the low-temperature
resistivity \cite{deHaasKondo1934} in gold wires containing a low
concentration of Cu impurities, it has also been realized by depositing
magnetic ad-atoms \cite{Manoharan2000,AgamSchiller2001,FieteEtAl2001}
or molecules on metallic surfaces.  Scanning tunneling microscopes
(STM) allow to manipulate and detect magnetic adatoms on metallic
substrates using the Kondo resonance \cite{Kondo1964,Kondo2005}.
Quantum corrals have been  built and a coherent interference of the electrons
was detected on the substrate surfaces using
STM \cite{Manoharan2000,AgamSchiller2001,FieteEtAl2001}. In 1998, David
Goldhaber-Gordon demonstrated in a seminal
paper \cite{NatureGoldhaberGordon1998} that the Kondo effect can also
be observed in single-electron transistors \cite{KastnerSET1992}
realized by a semiconductor quantum dot.  Both types of experiments
have opened up a new field of observing the Kondo-effect in
nanodevices \cite{Park2000,YuNatelson2004,*NatelsonCoII2005,TemirovLassieAndersTautz2008}.

While the equilibrium properties of the Kondo problem are
theoretically well understood by the virtue of Wilson's numerical
renormalization group  (NRG) \cite{Wilson75,BullaCostiPruschke2008}
approach and the exact Bethe ansatz solution \cite{Schlottmann89}, its
nonequilibrium properties are subject to active
research
\cite{Costi97,NordlanderEtAl1999,paaskeNonEqKondoMagField04,AndersSchiller2005,*AndersSchiller2006,Kehrein2005,SchiroFabrizio2009,Werner09,Schoeller2009a,*RT-RG-Strong-coupling,*schurichtDynCorrNeqKM09,Medvedyeva2013}.

In this paper, we address the fundamental question how the spatial
Kondo correlations are building up in time assuming an initially decoupled
impurity spin $\vec{S}_{\rm imp}$. We have investigated the
time-dependent spin-correlation function $\chi(\vec{r},t) =
\expect{\vec{S}_{\rm imp} \vec{s}(\vec{r})}(t)$
using the time-dependent NRG (TD-NRG) \cite{AndersSchiller2005,AndersSchiller2006}.
 $\vec{s}(\vec{r})$ denotes the spin density of the conduction electrons
at distance $R=|\vec{r}|$ from the impurity after switching on the
Kondo coupling at time $t=0$. Since the spins are initially
uncorrelated, this correlation function vanishes for $t\le 0$ and is a
measure of the building up of the spatial entanglement between the local
spin and the conduction-electron spin density.

For infinitely long times, the equilibrium spatial correlation
function of the Kondo model must be recovered. This correlation
function $\chi_\infty(\vec{r})=\lim_{t\to\infty} \chi(\vec{r},t) $ has
been investigated intensively by Affleck and
collaborators \cite{Barzykin1998,Affleck2001,Affleck2005,Affleck2008}
using field theoretical methods in the last 15 years. It accounts for
alternating ferromagnetic and antiferromagnetic correlations as
expected by spin-correlations mediated by the RKKY
mechanism. Furthermore, the crossover between different power-law
decays for short and long distances, found \cite{Borda2007,Holzner2009,Buessner2010,Mitchell2011} in
$\chi_\infty(\vec{r})$, occurs at a characteristic length scale $\xi_K
= v_F/T_K$, which has been interpreted as the size of the Kondo screening
cloud \cite{Barzykin1996,Barzykin1998,Affleck2001,Affleck2005,Affleck2008},
where $v_F$ is the Fermi velocity, and $T_K$ the Kondo temperature
governing the crossover from a free impurity spin at high temperatures
to the singlet formation for $T\to 0$.  The modulus of equilibrium correlation function 
$\chi_\infty(\vec{r})$ has also been investigated using a real-space DMRG  
\cite{Holzner2009}. Therefore, the two limits,
$t=0$ and $t=\infty$, are known and used as reference points for our
calculations.

The  time-dependent spin-correlation function
$\chi(\vec{r},t)$ contains the information about how the spin
correlations propagate through the system.  We have found 
for an antiferromagnetic Kondo coupling
(i) ferromagnetic correlations  propagating away from the impurity 
with the Fermi-velocity
$v_F$,  which defines  the ``light cone'' \cite{LiebRobinsonBound1972} of the problem,
and (ii) in addition finite and nonexponential small correlations
outside of this light cone. We have been able (iii)  to trace back the
correlations
outside of the light cone to the
intrinsic entanglement of the Fermi sea using a controlled
second-order expansion in the Kondo coupling constant and comparing 
the perturbative results with the full TD-NRG simulation. Since
$\chi(\vec{r},t)$ is not a response function, correlations outside of
the light cone are allowed.  

Any response
function, however, describing the transmission of signals  vanishes outside of
the light cone in accordance with relativity
if the momentum cutoff, is sent to infinity \cite{Medvedyeva2013}. For a finite
momentum cutoff this statement is weaken to fast decay on a length scale set 
by the inverse momentum cutoff.  At $T=0$, an algebraic decay is found in accordance with a broadened 
$\delta$-distribution function.

Our TD-NRG results agree remarkably well with our perturbative theory
for short and intermediate time and length scales and approach the
correct equilibrium correlation functions in the long-time limit.  Our
data confirm the recent findings by Medvedyeva {\it et al.}\ 
\cite{Medvedyeva2013} but also considerably extent their work: we
include the full spatial dependence that allows us to access the full
$2k_F$ oscillations inherent to the RKKY mediated correlations.
Furthermore, the crossover between short and long distances,
i.~e.,\ $R\ll \xi_K$ and $R\gg \xi_K$, including the Kondo physics at
low temperature in the strong coupling regime is fully accessible by the NRG,
which cannot be revealed by perturbative approaches.

Borda has pioneered the calculation of equilibrium spatial correlation
function for the Kondo model using the NRG \cite{Borda2007}. He has 
realized that this problem is equivalent to the two-impurity Kondo
model \cite{Jayaprakash1981,Jones_et_al_1987,Jones_et_al_1988} where
the second impurity spin has been removed while the local conduction
electron density operator $\vec{s}(\vec{r})$ is used as a probe for
the spin correlations.  By mapping the problem onto two
$\vec{r}$-dependent linear combinations of conduction electrons with
even and odd symmetry under spatial inversion around the midpoint
$\vec{r}/2$, the calculation of spatial correlations become accessible
to the NRG.  Thereby, the spatial
information \cite{Jones_et_al_1987,Jones_et_al_1988,AffleckLudwigJones1995,Borda2007}
is encoded in the energy dependent density of states (DOS) of the even
and odd bands. Since a single two-band NRG run is required for each distance $R$,
the numerical calculations are very involved and require an
independent NRG calculation with individually adapted bands for each
distance $R$.

An alternative approach for obtaining  real space information using a conventional single
site NRG has recently proposed  by Mitchell \emph{et al.}~\cite{Mitchell2011} who have
applied the exact equation of motion to relate specific $R$-dependent properties such
as the conduction electron density variation in the host to impurity properties using the free conduction
electron Green function. This approach, however, is not applicable to the 
spatially dependent spin-spin correlation function or the retarded spatial spin-spin susceptibility 
as investigate here.

In this paper, we use an improved mapping compared to Borda's original
work \cite{Borda2007}. Our modifications are able (i) to accurately
reproduce the analytically known
sum rules \cite{Barzykin1996,Barzykin1998,Borda2007} for the
spin-correlation function in the ferromagnetic and antiferromagnetic
Kondo regimes, (ii) reproduce the analytical spin-spin correlation
function of the decoupled Fermi sea exactly, and (iii) obtain sign
changes $\chi_\infty(\vec{r})$ at short and intermediate distances
expected from RKKY mediated correlations.  While Borda
reported \cite{Borda2007} that $\chi_\infty(\vec{r})$ remains negative
for all distances and Kondo couplings as can be seen in Fig.\ 2 of
Ref.\ \onlinecite{Borda2007}, we find oscillating and power-law
decaying $\chi_\infty(\vec{r})<0$ only for distances $R\gg \xi_K$ in
accordance with previous analytic predictions using a 1D field
theoretical approach \cite{ Barzykin1996,Barzykin1998}.  At short
distances, the Kondo screening is incomplete and, therefore,
alternating signs are found in $\chi_\infty(\vec{r})$.

We also discuss the spectral functions of the retarded spin-spin susceptibility 
as a function of $R$ and use these results to calculate the linear response
of the host spin density at a distance $R$ to a local magnetic field
applied on the impurity spin at the origin. Here, the response outside of the light cone
is suppressed. We benchmark the quality of the NRG spectral function with
the spin susceptibility of the metallic host without impurity for which the susceptibility 
can be calculate analytically.

\subsection{Plan of the paper} 

The paper is organised as follows.  We
begin with the definition of the model in Sec.\
\ref{sec:definition-of-the-model} and derive  the mapping to the
two-impurity model in Sec.\ \ref{sec:spatial-correlation}. After that,
we discuss in Sec.\ \ref{sec:sum-rules} the exact sum rules of the
equilibrium spin-correlation function for the Kondo regime and the
local moment regime relevant for a ferromagnetic Kondo coupling.

In Sec. \ref{sec:equilibrium}, we present our equilibrium results for
$\chi_{\infty}(R)$ for ferromagnetic and antiferromagnetic couplings
which differ slightly from the previously reported data by
Borda \cite{Borda2007} and discuss the effect of spatial dimensions.

In Sec. \ref{Sec:TD-NRG}, we provide a short summary of the TD-NRG
employed in the following to obtain the non-equilibrium quench
dynamics.  Section \ref{sec:TD-NRG-Kondo-regime} is devoted to our
numerical TD-NRG data on the temporal buildup of the spatial Kondo
correlations. In order to understand the surprising buildup of the Kondo
correlations outside of the light cone, a second-order perturbative
calculation presented in Sec.\ \ref{sec:perturbation-theory} is able
to provide an analytical explanation of the origin of this unusual
correlations found in the TD-NRG.  We use these analytical results to
also explain the spatial and temporal building up of spin-correlation for
ferromagnetic couplings in Sec.\ \ref{sec:LM-TD-NRG}.
In Sec. \ref{sec:finite-temp}, we extend the discussion to finite temperatures.
Section \ref{sec:response-function} is devoted to the retarded 
spin-spin susceptibilities and the time-dependent 
response of the conduction electron spin density as a function of
an suddenly applied local magnetic field within the linear response theory.
We conclude with
a summary and outlook.

\section{Theory}

\subsection{Definition of the model}
\label{sec:definition-of-the-model}

We investigate the spatial and temporal correlations between the 
conduction electron spin density 
$\vec{s}(\vec{r})$ and a localized impurity  spin $\vec{S}_{\rm imp} $
coupled locally
to the metallic host via Kondo interaction \cite{Kondo1964,Kondo2005}.
For a generic system, we can neglect the details of the atomic wave function and expand the 
spin-density operator $\vec{s}(\vec{r})$
in plane waves \cite{Hewson93}
\begin{eqnarray}
\vec{s}(\r) &=& \frac{1}{2} \frac{1}{NV_u}\sum_{\sigma\sigma'}\sum_{\k\k'} 
c^\dagger_{\k\sigma} [\vec{\mat{\sigma}}]_{\sigma\sigma'}  c_{\k'\sigma'} e^{i(\k'-\k)\r},
\end{eqnarray}
where $N$ is the number of unit cells in the volume $V$, $V_u=V/N$ is the volume of the unit cell,
and $\vec{k}$ the momentum vector. In an energy representation \cite{Wilson75,Hewson93},
the  Kondo Hamiltonian \cite{Kondo1964,Kondo2005} of the system,
\begin{eqnarray}
\label{eqn-Hk}
H &=& H_0 + H_K,
\\
H_0 &=& \sum_{\sigma} \int_{-D}^{D} d\e \, \e\, c^\dagger_{\e\sigma}c_{\e\sigma},
\nonumber \\
H_K &=& J \vec{S}_{\rm imp} \vec{s}_c(0),
\nonumber
\end{eqnarray}
describes this local impurity spin  located at the origin coupled via an effective Heisenberg coupling
to the unit-cell volume averaged conduction electron spin  \text{$ \vec{s}_c(\r) =V_u \vec{s}(\r)$} 
with a coupling constant $J$; $H_0$ accounts for the energy of the free conduction 
electrons.

Wilson \cite{Wilson75,BullaCostiPruschke2008} realized that the logarithmic 
divergencies generated by a perturbative
treatment \cite{Kondo1964,Kondo2005} of the Kondo coupling 
can be circumvented by discretizing the energy spectrum of the conduction 
band on a logarithmic grid \cite{Wilson75} using a dimensionless discretization
parameter $\Lambda>1$. Consequently, all intervals contribute equally to the divergence 
and the problem can be solved by iteratively adding these intervals with progressively 
smaller energy. 
The continuum limit is recovered for $\Lambda \to 1$. Using
an appropriate unitary transformation \cite{Wilson75}, the
Hamiltonian is mapped onto a semi-infinite chain, with the
impurity coupled to the first chain site. The $N$th link along the
chain represents an exponentially decreasing energy scale:
$D_N \sim \Lambda^{-N/2}$. Using this hierarchy of scales,
the sequence of finite-size Hamiltonians ${\cal H}_N$ for the
$N$-site chain is solved iteratively, discarding the high-energy states at
the conclusion of each step to maintain a manageable number
of states. The reduced basis set of ${\cal H}_N$ so obtained
is expected to faithfully describe the spectrum of the full
Hamiltonian on a scale of $D_N$, corresponding to the
temperature $T_N \sim D_N$. Details can be found in
the review \cite{BullaCostiPruschke2008} on the NRG by Bulla \emph{et al}.

\subsection{Spatial correlations} 
\label{sec:spatial-correlation}


While Wilson's original approach was tailored towards solving the
thermodynamics of  the local impurity and  the classifying the fixed points of
the Hamiltonian, we are explicitly interested in the  time evolution of the
spatial spin-correlation functions
$\expect{\vec{S}_{\rm imp}\vec{s}(\vec{r})}(t)$ at the distances $R$.
For a rotational invariant system considered here, this quantity is
isotropic and only dependent on $R$. The correlations between those
spatially well separated points are mediated by the conduction
electrons which is linked to the RKKY interaction.

Borda has realized \cite{Borda2007} that the calculation of the spatial
correlations  is related to a simplified two-impurity
problem. Originally Jones \emph{et al}.~\cite{Jones_et_al_1987,Jones_et_al_1988} have
extended the NRG \cite{Wilson75} to a two-impurity Kondo model, 
\begin{eqnarray}
\label{eqn:TI-KM}
H&=& H_0
+\sum_{i=\pm} J_i  \vec{S}^i_{\rm imp} \vec{s}_c(\vec{R_i})
\, ,
\end{eqnarray}
where the two impurity spins $\vec{S}^i_{\rm imp}$ are located at the position 
$\vec{R}_\pm= \pm \vec{r}/2$
and coupled to the same conduction band. The spatial dependence is included into
the two nonorthogonal energy-dependent field operators
\begin{eqnarray}
c_{\e\sigma,\pm } &=& \frac{1}{\sqrt{N\rho(\e)}} \sum_{\k} \delta(\e-\e_{\k}) c_{\k\sigma}e^{\pm i\vec{k}\vec{r}/2},
\end{eqnarray}
which are combined to even (e) and odd (o)
parity eigenstates \cite{Jayaprakash1981,Jones_et_al_1987,Jones_et_al_1988,AffleckLudwigJones1995} 
\begin{eqnarray}
c_{\e\sigma,e } &=&\frac{1}{ N_e(\e)} \left( c_{\e\sigma,+ } +c_{\e\sigma,-} \right),
\non
c_{\e\sigma,o } &=& \frac{1}{N_o(\e)} \left( c_{\e\sigma,+ } -c_{\e\sigma,-} \right),
\end{eqnarray}
of $H_0$.
The dimensionless normalization functions $N_{e(o)}(\e)$,
\begin{eqnarray}
N^2_{e}(\e)&= &
\frac{4}{N\rho(\e)} \sum_{\k} \delta(\e-\e_{\k}) \cos^2\left(\frac{\k \vec{r}}{2}\right),
 \non
N^2_{o}(\e)&=&
\frac{4}{N\rho(\e)} \sum_{\k} \delta(\e-\e_{\k}) \sin^2\left(\frac{\k \vec{r}}{2}\right), \label{EqNormFactor}
\label{eqn-N_eo}
\end{eqnarray}
are computed from the anticommutation relation 
$\{ c^{\phantom{\dagger}}_{\e\sigma,\alpha },c^\dagger_{\e\sigma,\alpha' }\}
=\delta(\e-\e')\delta_{\alpha\alpha'}\delta_{\sigma\sigma'}$.
Note that both densities $N_e(\e)$ and $N_o(\e)$ depend on the distance $R=|\vec{r}|$ 
and are not normalized. 
$\rho(\e)$  denotes the conduction band density of states of the original band.

The two-impurity Hamiltonian (\ref{eqn:TI-KM}) can be written in terms
of these even and odd fields and solved using the
NRG \cite{Jones_et_al_1987,Jones_et_al_1988}.
By  omitting one of the two-impurity spins \cite{Borda2007},
i.\ e.~, $\vec{S}^-_{\rm imp} $, the original Kondo Hamiltonian
(\ref{eqn-Hk}) is recovered for an impurity spin located at
$\vec{R}_{+}$. Furthermore, $\vec{s}_c(\vec{R_-})$ can be used for
probing the isotropic spin-correlation function
$\expect{\vec{S}_{\rm imp}\vec{s}(\vec{r})}$.

The local  even or odd parity conduction electron operators 
coupling to the impurity spin take the form
\begin{eqnarray}
f_{0\sigma,e(o)} &=& \frac{1}{\bar N_{e(o)}}\int d\e \sqrt{\rho(\e)} N_{e(o)}(\e) c_{\e\sigma,e(o)} 
\komma
\label{eqn:7}
\end{eqnarray}
and its anti-commutator $\{ f_{0\sigma,e(o)} ,f^\dagger_{0\sigma',e(o)}\} =\delta_{\sigma\sigma'}$
determines the dimensionless normalization constants
\begin{eqnarray}
\bar N_{e(o)} &=& \left[  
\int d\e N^2_{e(o)}(\e) \rho(\e)
\right]^{1/2} \label{N2(e)}
\punkt
\end{eqnarray}
These constants enter the definition of the effective parity density of states
\begin{eqnarray}
\label{eqn:9}
\rho_{e(o)}(\e) &=& \frac{1}{\bar N^2_{e(o)}} N^2_{e(o)}(\e)\rho(\e) \label{rho(e)}
\end{eqnarray}
which accounts for the position dependence and are used in the construction of the 
NRG tight-binding chain (for details see the NRG review in 
Ref.\ \cite{BullaCostiPruschke2008}).

Then, the original Kondo Hamiltonian (\ref{eqn-Hk}) is expanded in these orthogonal even and odd fields:
\begin{eqnarray}
\label{eqn:mapped-Hk}
H &=& 
 \sum_{\sigma} \sum_{\alpha=e,o} \int_{-D}^{D} d\e \, \e\, c^\dagger_{\e\sigma,\alpha}c_{\e\sigma,\alpha}
\nonumber
\\
&&\nonumber
+
\frac{J}{8}\sum_{\sigma\sigma'}
\left(\bar N_e f^\dagger_{0\sigma,e} + \bar N_o f^\dagger_{0\sigma,o} 
\right)
[\vec{\mat{\sigma}}]_{\sigma\sigma'}
\nonumber \\
&&
\phantom{\frac{J}{8}\sum_{\sigma\sigma'}}
\times
\left(\bar N_e f_{0\sigma,e} + \bar N_o f_{0\sigma,o} 
\right)\vec{S}_{\rm imp}
\end{eqnarray}
after positioning  the impurity spin at $\vec{R}_{+}$.  
The spin-density operator at $\vec{R}_{-}$  entering the spatial correlation function is
given by
\begin{eqnarray}
\vec{s}(\vec{R}_{-}) &=& \frac{1}{8V_u} \sum_{\sigma\sigma'} 
\left(\bar N_e f^\dagger_{0\sigma,e} - \bar N_o f^\dagger_{0\sigma,o} 
\right)
[\vec{\mat{\sigma}}]_{\sigma\sigma'}
\non
&&\times
\left(\bar N_e f_{0\sigma,e} - \bar N_o f_{0\sigma,o} 
\right),
\end{eqnarray}
where $V_u$ accounts for its dimensions.

Note the inclusion of the proper $R$-dependent normalization constants $\bar N_e,\bar N_o$
into the Hamiltonian and the spin-density operator $\vec{s}(\vec{R}_{-})$, which are crucial  for recovering
the exact sum rules discussed in the following section. 
Furthermore, we use properly normalised
conduction bands with an energy-dependent density of states as defined in Eq.\ \eqref{eqn:9}
and  renormalized Kondo couplings $J$ in Eq.\ \eqref{eqn:mapped-Hk}, while Borda
included an unnormalised DOS $\rho_i(\e) = N^2_{i}(\e)\rho(\e)$ into the kinetic energy term, see Eq.\ (6) in Ref.\ \cite{Borda2007}.

\subsection{Sum-rule of the spatial correlation function}
\label{sec:sum-rules}

The quality of the calculated spatial correlation function 
can be verified by a sum-rule derived for the strong-coupling fixed point at $T=0$.
Since the ground state $\ket{0}$ is a singlet  in the Kondo regime, 
the application of the total spin operator of the system
comprising local and total conduction electron spin
\begin{eqnarray}
	\vec{S}_\mathrm{tot}\ket{0}= \left( \vec{S}_\mathrm{imp} + \int s(\vec{r}) \; d^Dr \right)\ket{0}=0 \end{eqnarray}
must vanish. Consequently, the correlator $\bra{0} \vec{S}_\mathrm{imp}\vec{S}_\mathrm{tot} \ket{0}$
also  vanishes, 
\begin{eqnarray}
	\bra{0} \vec{S}_\mathrm{imp}\vec{S}_\mathrm{tot} \ket{0}=\frac{3}{4} + \int \bra{0} \vec{S}_\mathrm{imp} \vec{s}(\vec{r}) \ket{0} \; d^Dr =0 
	\, ,
\end{eqnarray}
and, therefore, $\chi_\infty(r)$ must obey the sum rule:
\begin{eqnarray}
	\int \langle \vec{S}_\mathrm{imp} \vec{s}(\vec{r}) \rangle \; d^Dr = -\frac{3}{4} \punkt 
	\label{sum rule}
\end{eqnarray}
at $T=0$.  The spin-correlation function is isotrop for a generic system.
Substituting the  
dimensionless variable $x=k_\mathrm{F} R/\pi$ and angular integration  yields
\begin{eqnarray}
	\frac{C_D \pi^D}{k^D_\mathrm{F}}\int_{0}^\infty x^{D-1} \chi_\infty(x) \; dx =-\frac{3}{4}
	\, ,
	\label{eqn:sum-rule-dimmless}
\end{eqnarray}
where $D$ is the dimension, $C_1=2$, $C_2=2\pi$ and \text{$C_3=4\pi$}.

For the linear dispersion $\epsilon(|\vec{k}|)=v_\mathrm{F}\left(
|\vec{k}|-k_\mathrm{F} \right)$ the Fermi wave vector in the different
dimensions is given by \text{$k_\mathrm{F}=\pi/2V_u$} in 1D,
$k_\mathrm{F}=\sqrt{\pi/V_u}$ in 2D and
\text{$k_\mathrm{F}=\left(\pi^2/V_u\right)^{1/3}$} 
in 3D.  The volume of a
unit-cell $V_u$ in the Fermi wave vector is canceled by the factor
$1/V_u$ in $\vec{s}(x)$ after substituting $\vec{s}_c(\r) =V_u
\vec{s}(\r)$ in the correlation function.

Numerically evaluating the sum rule (\ref{eqn:sum-rule-dimmless}) using
our NRG correlation function, we have confirmed the theoretical value
of $-\frac{3}{4}$ with an error less than $2\%$ in 1D.  Since
$\chi_\infty(R)\propto R^{-(D+1)}$ for $R\to\infty$,  the
integral kernel $ R^{D+1}\chi_\infty(R)$
is very susceptible to numerical errors in higher
dimensions. Therefore, the accuracy decreases with increasing
dimensions.  For distances $k_\mathrm{F} R/\xi_\mathrm{K}\gg1$ a very
high number of kept states in the NRG calculation are required to
prevent the integral $\int_{0}^\infty R^{D-1} \chi_\infty(R) \; dR$
from diverging.

For a ferromagnetic coupling the Hamiltonian approaches the
local moment (LM) fixed point with a decoupled impurity spin. Using the
same argument as above yields the sum-rule
\begin{eqnarray} \int \langle \vec{S}_\mathrm{imp} \vec{s}(\vec{r})
\rangle \; d^Dr=0 \punkt
	\label{eq:LM-sum rule}
\end{eqnarray} valid for $J<0$. Consequently, we expect an oscillatory
solution for $\chi_\infty(R)$ with sign changes at all length scales
and a decay $R^{-\alpha}$ where $\alpha\ge D$: the spin correlation
function will be significantly different in the ferromagnetic and in
the antiferromagnetic regime.

\subsection{Effective densities of states in 1D, 2D and 3D}

\begin{figure}[t]
\centering
\includegraphics[width=0.5\textwidth]{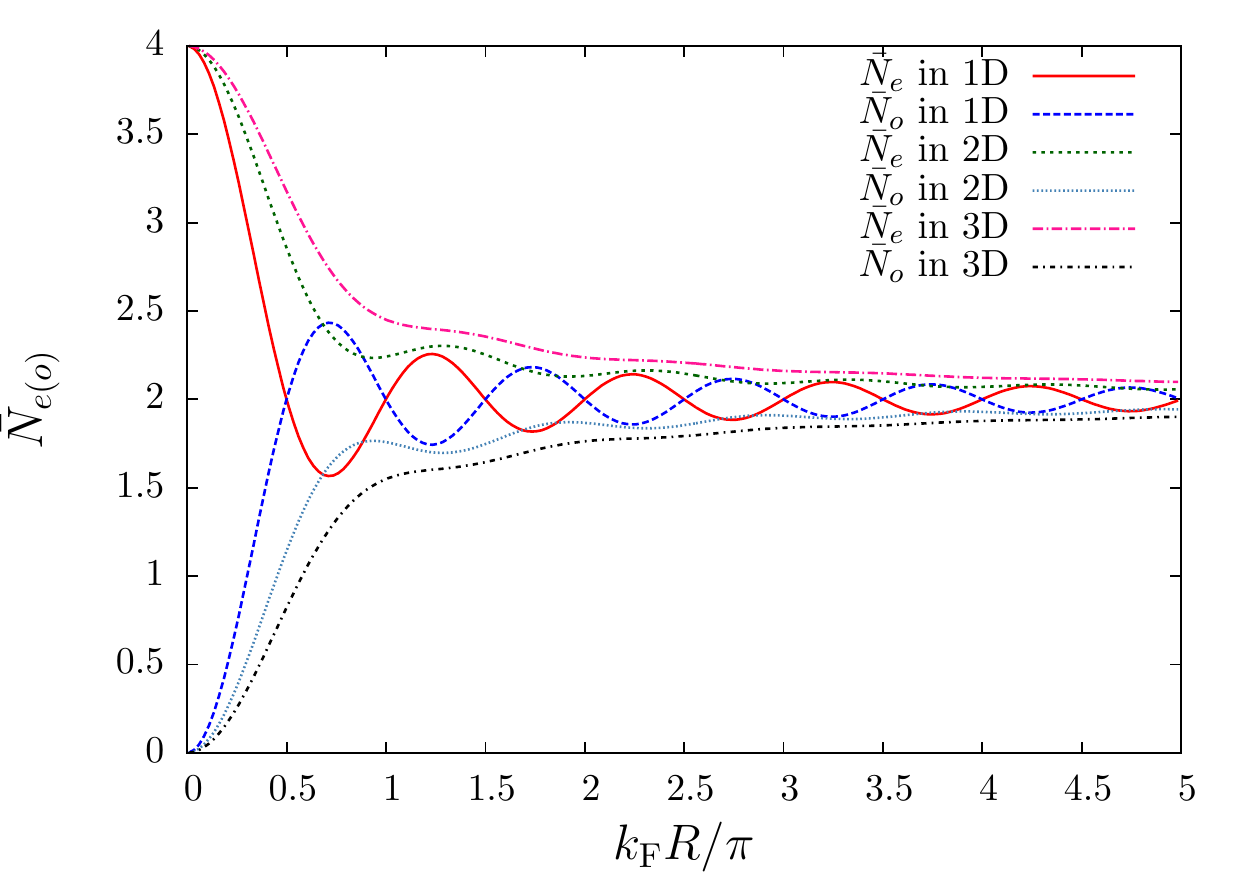}
\caption{(Color online) Normalization constants $\bar N_{e(o)}$ for different dimensions $D$
vs the dimensionless distance $x=k_\mathrm{F}R/\pi$. 
For $R\rightarrow \infty$ $\bar N_{e}$ is equal to $\bar N_{o}$.}
\label{NormFactors}
\end{figure}

The spatial correlations depend on the dimensionality of the host. For a given distance $R$, 
the dimensionality enters primarily via the dimension of the wave vector $\k$ in the 
Eqs.\ (\ref{eqn-N_eo}) and the energy dispersion $\e_{\k}$ of the host. 

In order to obtain information on generic spectral densities 
$N^2_{e(o)}(\epsilon)\rho(\epsilon)$ appearing in  the Eqs.~\eqref{N2(e)} and \eqref{rho(e)} 
we assume a isotropic linear dispersion
$\epsilon_{\k} =v_\mathrm{F}\left( |\vec{k}|-k_\mathrm{F} \right)$,
where $v_\mathrm{F}$ is the Fermi velocity  and $k_\mathrm{F}$ the Fermi wave-vector. 
Inserting the dispersion in equations \eqref{EqNormFactor} yields in 1D to
\begin{eqnarray}
	N^2_{e(o)}(\epsilon)\rho(\epsilon)
	=2\rho_0\left[ 1 \pm \cos\left( x\pi \left(1+\frac{\epsilon}{D}\right) \right) \right] \label{eq:Norm_1D}
\end{eqnarray}
where $\rho_0=1/2D$ is the constant density of states and \text{$x=k_\mathrm{F}R/\pi$}.
In higher dimensions we perform the angular integration to obtain for 2D
\begin{eqnarray}
	N^2_{e(o)}(\epsilon)\rho(\epsilon)=2\rho_0\left[ 1 \pm J_0\left( x \pi \left( 1+\frac{\epsilon}{D}\right) \right) \right] \label{eq:Norm_2D}
\end{eqnarray}
with the zeroth Bessel function $J_0(x)$. In 3D, the
effective densities of states \cite{Jones_et_al_1987,Jones_et_al_1988}  reads
\begin{eqnarray}
	N^2_{e(o)}(\epsilon)\rho(\epsilon)=2\rho_0\left[ 1 \pm \frac{\sin\left( x \pi\left( 1+\frac{\epsilon}{D} \right) \right)}{x \pi\left( 1+\frac{\epsilon}{D} \right)} \right] \punkt \label{eq:Norm_3D}
\end{eqnarray}
Note that in 2D and 3D $\rho(\epsilon)$ is not constant for a linear dispersion,  and 
$\rho(\epsilon)=\rho_0=1/2D$ is a simplification. 

The normalization constants $\bar N_{e(o)}$ reveal important
information on the admixture of even and odd bands for a given
distance $R$. They are shown as a function of the dimensionless distance
$x=k_\mathrm{F}R/\pi$ for different dimensions D in
Fig.~\ref{NormFactors}.  Clearly, $\bar N_{o}(x=0)=0$ in any
dimension: the odd band decouples from the problem, and the standard
Kondo model is recovered which allows to calculate local ($R=0$)
expectation values within the standard single band NRG \footnote{In order to avoid different numerical
accuracy for $R=0$ and $R>0$ calculations, we have used $k_{\rm F}R/\pi=0.01$
in the NRG calculations for $R\to 0$.}.

With increasing $R$, the oscillations of the even and odd density of
states $\rho_{e(o)}$ decay as $\propto R^{(1-D)/2}$.  For large
distances, the even and odd bands become equal and the normalization
constants approach the same value. For 1D, strong oscillations are
observed for short distances which are suppressed in higher
dimensions. Apparently the $R$ dependence will be more pronounced in
lower dimensions and the correlation function will decay with the
different power law than in higher dimensions.

\section{Equilibrium physics:  spatial correlation}

\label{sec:equilibrium}

\subsection{Kondo regime: Short distance vs large distance properties}

\begin{figure}[t]
\centering
 \flushleft{(a)}
 \includegraphics[width=0.5\textwidth]{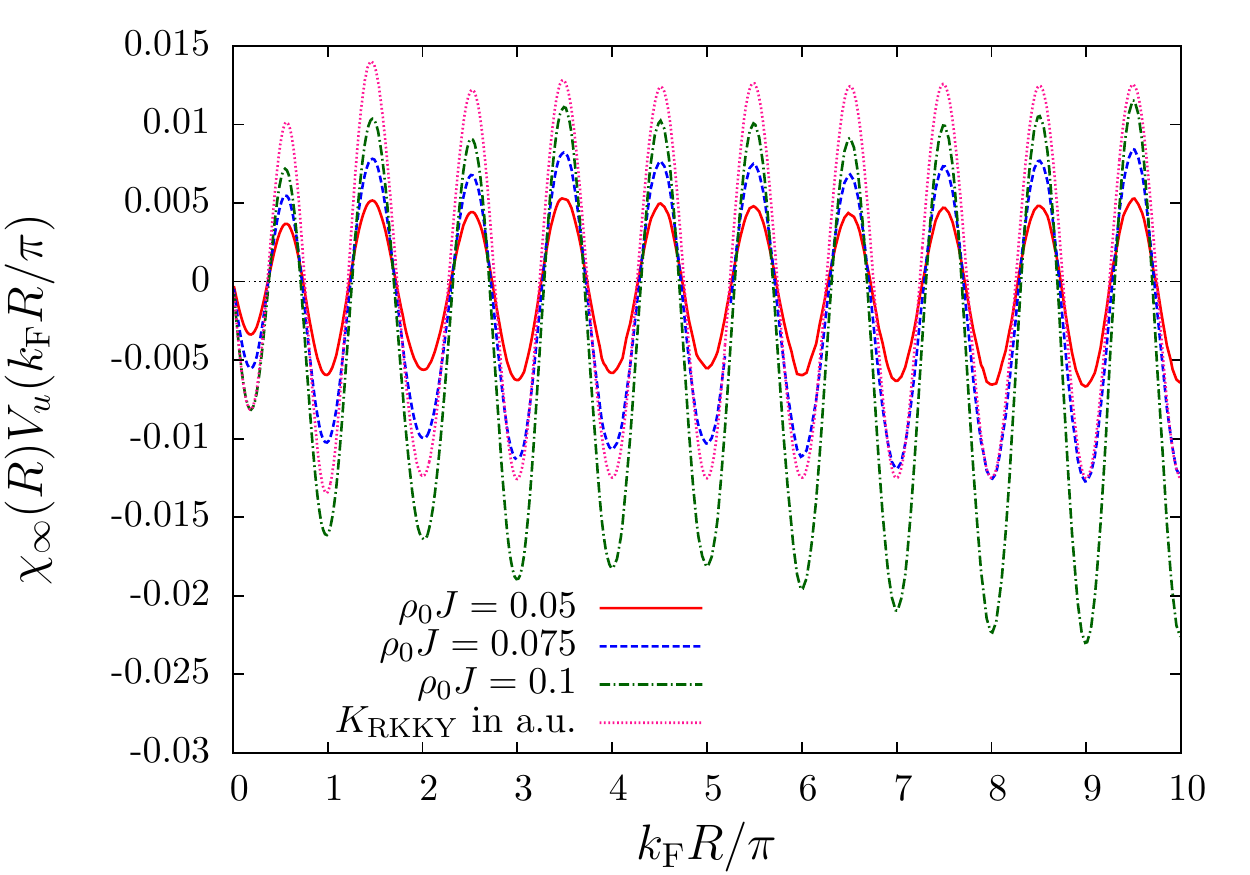}
 \flushleft{(b)}
 \includegraphics[width=0.5\textwidth]{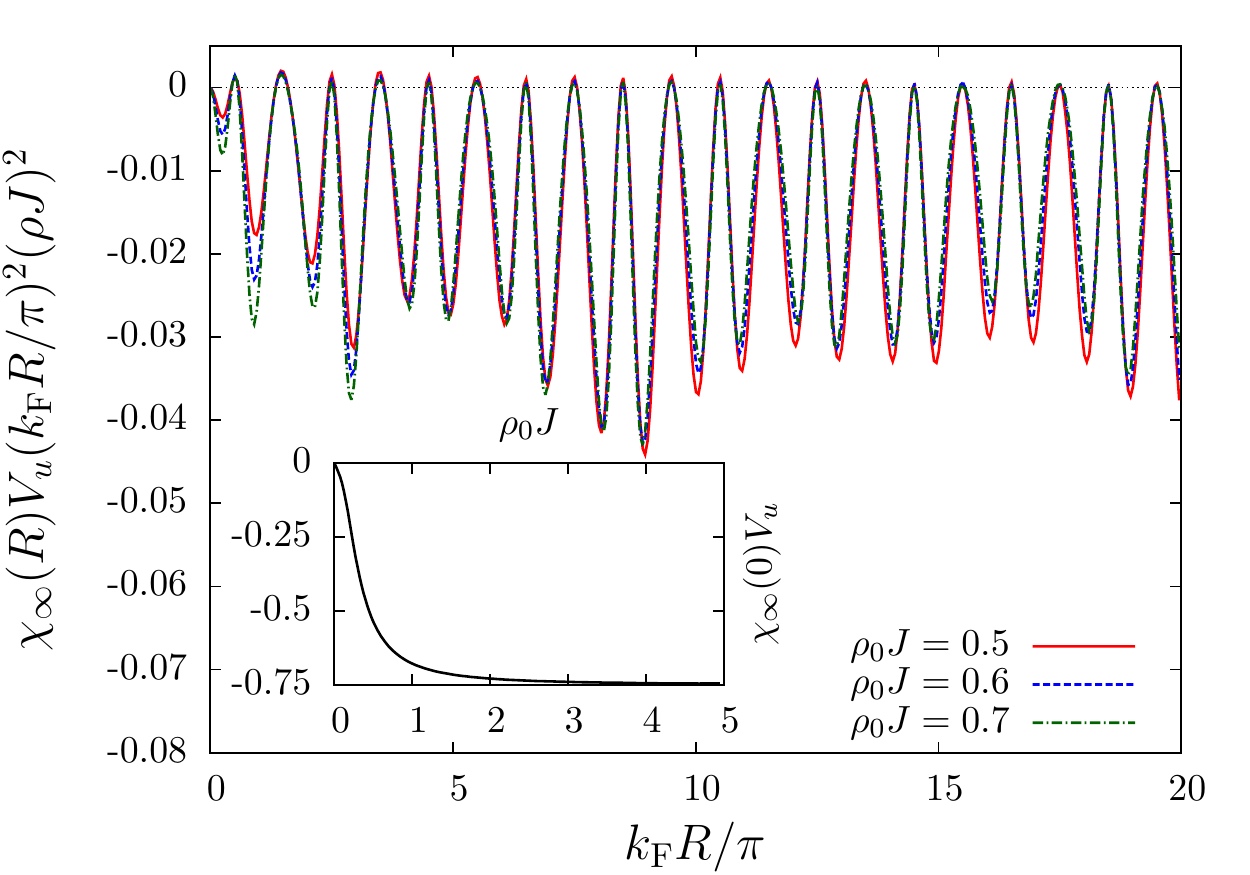}

\caption{(Color online) 
The spin-correlation function  (a)
$R\chi_{\infty}(R)= R\langle \vec{S}_\mathrm{imp} \vec{s}(\vec{r}) \rangle$ 
as a function of the dimensionless distance $x=k_F R/\pi$ 
in 1D for small Kondo couplings $\rho_0 J=0.05,0.075,0.1$
and $R/\xi_K\ll 1$. We added 
RKKY interaction between the impurity spin and a probe
spin in distance $R$ for comparison.
(b) 
$R^2 \chi_{\infty}(R)$ as a function of the dimensionless distance $x$ 
in 1D for larger Kondo couplings $\rho_0 J=0.5,0.6,0.7$.
The inset shows the value of the correlation function at the origin $\chi_\infty(0)$ vs $\rho J$. 
For large $J$, the value $-3/4$ is reached.
}

\label{fig:2ab}
\end{figure}

\begin{table}[t]
 \begin{tabular}{c|c|c}
  $\rho_0 J$  & $T_\mathrm{K}/D$  & $k_\mathrm{F} \xi_\mathrm{K}$ \\ \hline \hline
  $0.05$  & $1\cdot 10^{-10}$  & $ 1\cdot 10^{10}$ \\ \hline
  $0.075$  & $1.7\cdot 10^{-7}$  & $5.88 \cdot 10^{6}$ \\ \hline
  $0.1$  & $7.5\cdot 10^{-6}$  & $1.33 \cdot 10^{5}$ \\ \hline
  $0.15$  & $3.9\cdot 10^{-4}$  & $2564.10$ \\ \hline
  $0.3$  & $0.0204$  & $49.02$ \\ \hline
  $0.5$  & $0.0749$  & $13.35$ \\ \hline
  $0.6$  & $0.115$  & $8.69$ \\ \hline
  $0.7$  & $0.2103$  & $4.76$ 
 \end{tabular}
  \caption{The Kondo temperature $T_K$ and length scale $\xi_\mathrm{K}$ for different Kondo couplings.
  The Kondo temperatures have been obtained from the NRG level flow - see text.}
  \label{tab:Kondo}
\end{table}

There are two characteristic length scales in the problem: $1/k_F$
defined by the metallic host, governing the power-law decay of
$\chi_\infty(R)$ and its RKKY oscillations, and the Kondo length scale $\xi_K$, sometimes
referred to as size of the Kondo screening
cloud \cite{Barzykin1996,Barzykin1998,Affleck2001,Affleck2005,Affleck2008}.
Since $\xi_K$ increases exponentially with decreasing $J$, we use
different $J$ to present data for the two different regimes $R<\xi_K$
and $\xi_K<R$.  The results for $\chi_\infty(R)$ for these two different
regimes are shown in Fig. \ref{fig:2ab}.

In our calculations, the Kondo temperature $T_\mathrm{K}$ has been defined from
the NRG level flow: $T_\mathrm{K}$ is the energy scale 
at which the first excitation reaches $80$ \% of its fixed point value.
The Kondo temperature $T_\mathrm{K}$ and length scale $\xi_\mathrm{K}$ 
for different Kondo couplings
are stated in Table. \ref{tab:Kondo}. All correlation
functions have been obtained for $T/T_\mathrm{K} \to 0$. The sum rule
(\ref{eqn:sum-rule-dimmless}) of $\chi_\infty(R)$ is numerically
fulfilled up to typically 2\% error in 1D.

In contrast to the original work by Borda \cite{Borda2007}, we observe
ferromagnetic and anti-ferromagnetic correlations for short distances
in accordance with 
predictions \cite{Barzykin1996,Barzykin1998,Affleck2001,Affleck2005,Affleck2008}
by Affleck and his collaborators as presented in Fig.\ \ref{fig:2ab}(a).  
For distances $R\ll \xi_\mathrm{K}$, the impurity is still unscreened
and impurity spin behaves more like a free spin. 
We have plotted the rescaled correlation function $R\chi_\infty(R)$
to reveal the  $1/R$ decay at short distances in 1D \cite{Borda2007} 
stemming from the  analytical form of the RKKY interaction.

We  have  also added the
RKKY interaction $K_\mathrm{RKKY}$ between the impurity spin and a
fictitious probe impurity spin  at distance $R$  obtained in second order
perturbation theory for comparison (details of the calculations
can be found in Appendix \ref{sec:appendix-RKKY}).
The
oscillating part of $\chi_\infty(R)$, and the positions of the minima
and maxima nicely agree with the RKKY interaction $\propto
\cos(2k_\mathrm{F} R)$.  For multiple integers 
$x = k_\mathrm{F}R/\pi =n$ the correlation function has minima and for odd
multiple $x= n+ 1/2$ a maxima is found.

In order to access larger distances $R\gg\xi_K$, we increase $\rho_0
J$.  The rescaled spin correlation function $R^2\chi_{\infty}(R)$
depicted in Fig.\ \ref{fig:2ab}(b) clearly reveals the power-law decay of the envelop
function at distances $R/\xi_K\gg 1$ as $1/R^2$.
Furthermore, we only find antiferromagnetic correlations for $R/\xi_K\gg 1$, and
$\chi_{\infty}(R)$ remains negative at all distances.  In this regime, the maxima 
have the value $\chi_\infty(R)=0$ as noted earlier 
\cite{Barzykin1996,Barzykin1998,Affleck2001,Affleck2005,Affleck2008}:
The impurity spin is screened by the conduction
band electrons and the envelope of $\chi_\infty(R)$ has to decreases
faster.  

We observe this crossover from 
a $1/R$ to a $1/R^2$ decay at around $R \approx
\xi_\mathrm{K}$.   We also find
universal behavior for the envelope of $\chi_\infty(R)$ for distances
$k_\mathrm{F}R/\pi>1$ and can reproduce Fig.\ 3 of Ref.\
\cite{Borda2007} (not shown here.)

Since we have plotted  $R^2\chi_{\infty}(R)$ which vanishes for $R=0$, 
the information of $\chi_{\infty}(0)$  is  not included in the main panels of 
Fig.\ \ref{fig:2ab}. 
Therefore, we have added the $\rho_0 J$ dependence
of the local spin-correlation function $\chi_{\infty}(R=0)$ as inset to Fig.\ \ref{fig:2ab}(b). 
$\chi_{\infty}(0)<0$, as expected for a local antiferromagnetic coupling, and 
the strong coupling value of $-3/4$ is approached for large $J$: 
almost the whole contribution to the sum rule
\eqref{eqn:sum-rule-dimmless} is located in the first antiferromagnetic peak
at $R=0$, and $\chi_\infty(R)$ has to decay very rapidly with increasing $R$.

\begin{figure}[t]
\centering

\includegraphics[width=0.5\textwidth]{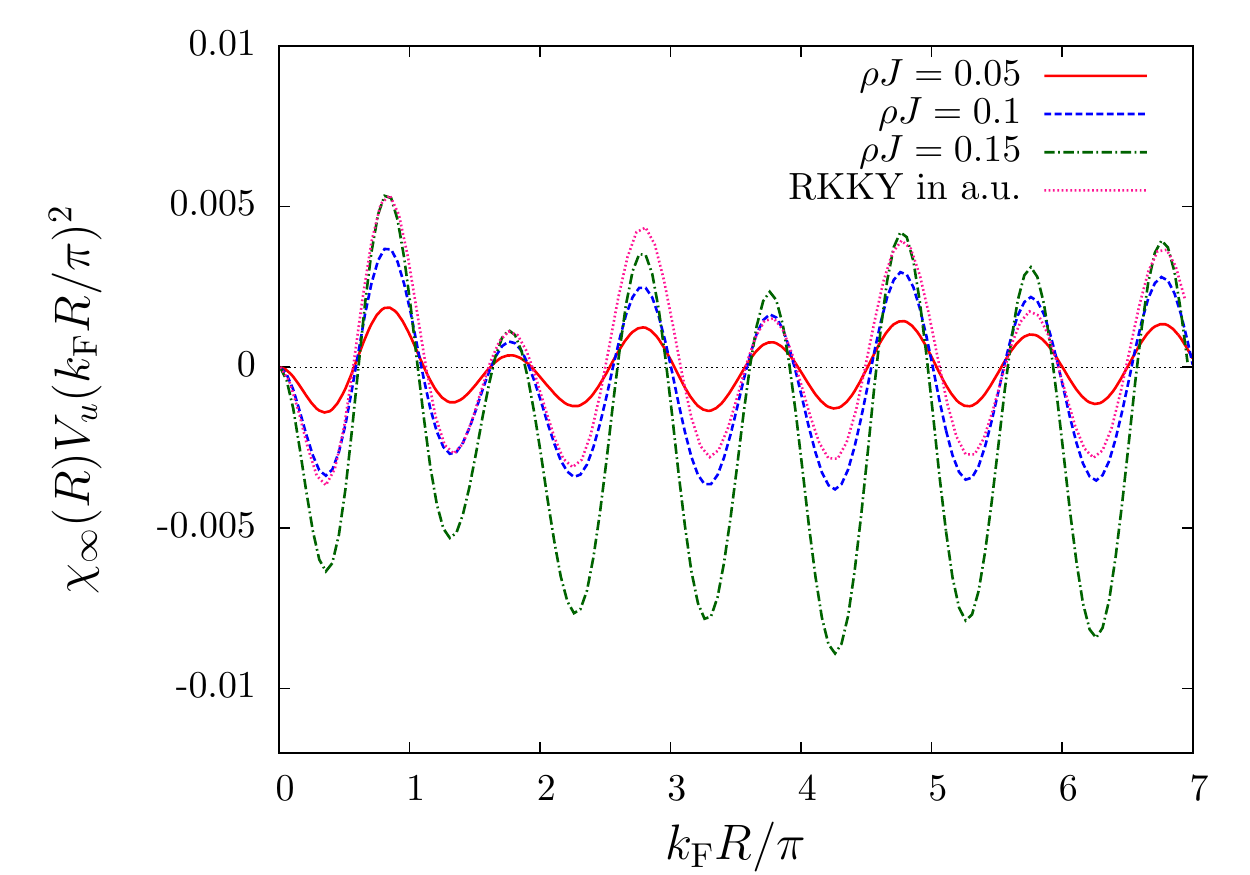}

\caption{(Color online) 
The spin-correlation function $R^2\chi_{\infty}(R)$ 
as a function of the dimensionless distance $x=k_F R/\pi$ 
in 2D. In higher dimensions, the envelope of the RKKY and $\chi_{\infty}(R)$ has a more complicated structure. 
Every second maximum has a lower amplitude.
NRG parameters are $\Lambda=5$ and $N_s=3000$.
}

\label{2D-Verlauf}
\end{figure}

In Fig.\ \ref{2D-Verlauf}, the short distance behavior of
$\chi_\infty(R)$ in 2D is shown.  As for 1D case the
oscillating part and the positions of the minima and maxima of
$\chi_\infty(R)$ and the 2D RKKY-interaction nicely agree.  In
contrast to 1D, the RKKY interaction acquires a more
complex mathematical structure even for a simplified linear 
dispersion replacing the  simple $\cos(2 k_\mathrm{F}R)$ oscillations in 1D.
Therefore, modification to the $\cos(2 k_\mathrm{F}R)$ behavior
must be taken into account when analyzing experimental data.

\subsection{Ferromagnetic couplings $J<0$ in 1D}
\label{sec:ferro-J-equ}

\begin{figure}[t]
\centering

\includegraphics[width=0.5\textwidth]{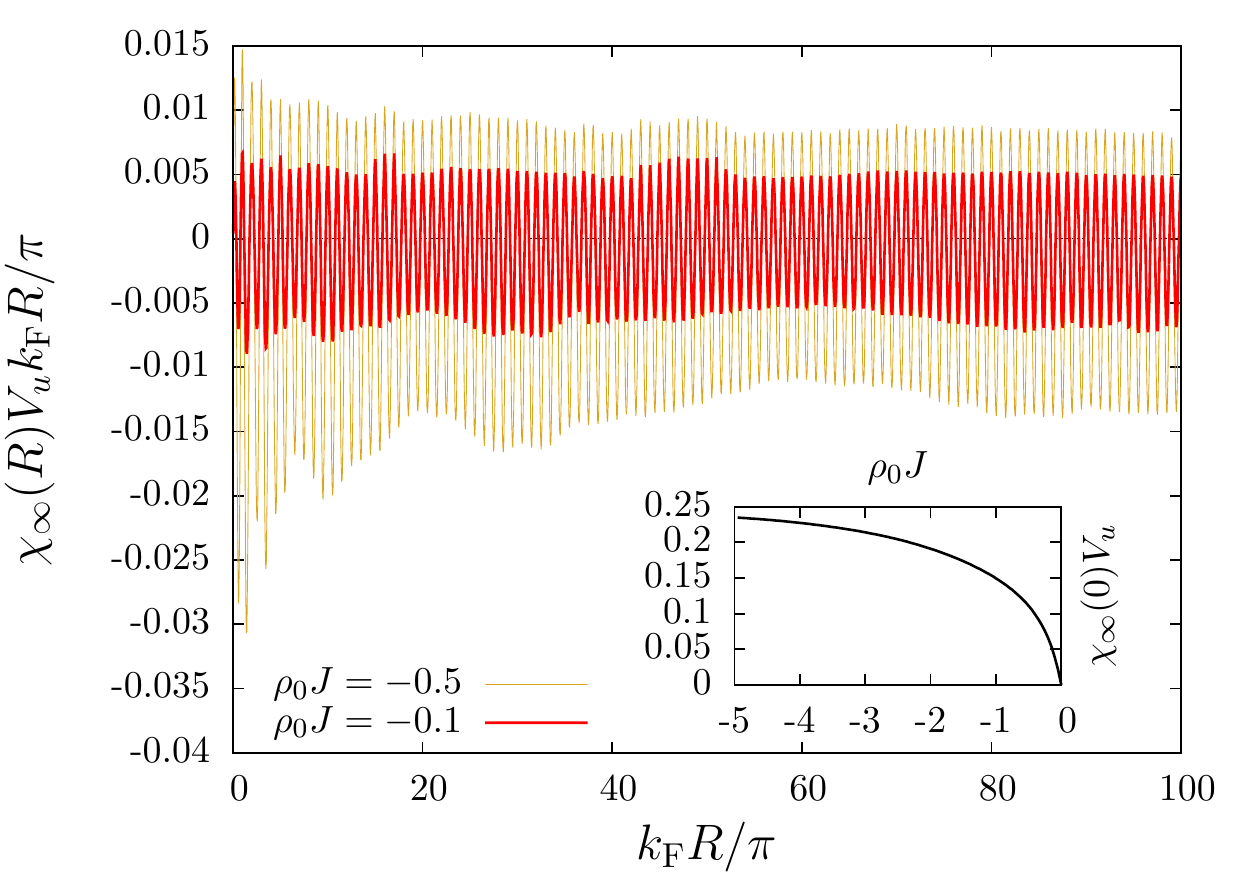}

\caption{(Color online) 
The spin-correlation function $R\chi_{\infty}(R)$ 
as a function of the dimensionless distance $x=k_{\rm F} R/\pi$ 
in 1D for two different ferromagnetic Kondo couplings \text{$\rho_0 J=-0.1$} and $\rho_0 J=-0.5$ for $T\to 0$.
The inset shows $\chi_\infty(0)$ vs $\rho J$. For large ferromagnetic couplings $0.25$ is reached.
NRG parameters are $\Lambda=3$ and $N_s=1400$.
}

\label{fig:ferro-chi-equi}
\end{figure}

Up until now, we have focused only on antiferromagnetic Kondo coupling
causing the Kondo-singlet formation described by the
strong-coupling (SC) fixed point  for $T\to 0$. We extend our discussion of the
equilibrium correlation function to the ferromagnetic regime
characterized by the local-moment (LM) fixed point and a twofold degenerate
ground state.  As pointed out above, the sum rule for
$\chi_{\infty}(\vec{r})$ predicts that the spatial integral of the
correlation function vanishes.  For a ferromagnetic coupling, $\xi_{\rm
K}\to \infty$, and the correlation function is also oscillating as $\cos(2k_{\rm F} R)$ 
and  decaying as $1/R$ for at all distances in 1D.

Exemplifying our findings, we depicted $R\chi_{\infty}(R)$ for 
two different Kondo coupling, $\rho_0
J=-0.1$ and $\rho_0 J=-0.5$,  in Fig.\ \ref{fig:ferro-chi-equi}.  We
numerically checked the sum rule and found deviations from zero by
less than 1\% in 1D.  The RKKY oscillations and $1/R$ decay of the
envelop function are clearly visible up very larger distances  
$k_{\rm F} R/\pi =100$.

In the ferromagnetic regime, the local  spin-correlation function $\chi_{\infty}(0)$ 
must be positive and approaches its upper limit of {$\chi_{\infty}(0)\to 1/4$ 
for $J\to -\infty$ as shown in the inset of Fig.\ \ref{fig:ferro-chi-equi}.
In order to fulfill the sum rule, $\chi_{\infty}(R)$ does not oscillate
symmetrically around the $x$ axis: $\chi_{\infty}(R)$ must be
slightly shifted to antiferromagnetic correlations to
compensate the  ferromagnetic peak at $R=0$.

\section{Nonequilibrium dynamics}

\subsection{Extension of the NRG to nonequilibrium: the TD-NRG}
\label{Sec:TD-NRG}

The TD-NRG has been
designed \cite{AndersSchiller2005,AndersSchiller2006} to track the
real-time dynamics of quantum-impurity systems following an abrupt
quantum quench such as considered here; at $t=0$, we switch on the
Kondo coupling $J$ between the prior decoupled impurity spin and the
metallic host.
 
Initially, the entire system is characterized
by the density operator  of the free electron gas
\begin{equation}
\hat{\rho}_0 =
     \frac{ e^{-\beta H_0 }}
          {{\rm Tr}
                \left[
                        e^{-\beta H_0 }
                \right] } ,
\label{rho-0}
\end{equation}
at time $t=0$
when the Kondo interaction $H_K$ is
suddenly switched on: $H= H_0 + H_K$.
The density operator evolves thereafter in
time according to
\begin{equation}
\hat{\rho}(t > 0) =
    e^{-i t H} \hat{\rho}_0
    e^{i t H } .
\label{rho-of-t}
\end{equation}
Our objective is to use the NRG to compute the
time-dependent expectation value $O(t)$ of a
general local operator $\hat{O}$. As shown in
Refs~\cite{AndersSchiller2005,AndersSchiller2006},
the result can be written in the form
\begin{eqnarray}
\langle \hat{O} \rangle (t) &=&
        \sum_{m}^{N}\sum_{r,s}^{\rm trun} \;
        e^{i t (E_{r}^m - E_{s}^m)}
        O_{r,s}^m \rho^{\rm red}_{s,r}(m) ,
\label{eqn:time-evolution-intro} 
\end{eqnarray}
where $E_{r}^m$ and $E_{s}^m$ are the dimension-full
NRG eigenenergies of the perturbed Hamiltonian at
iteration $m \le N$, $O_{r,s}^m$ is the matrix
representation of $\hat{O}$ at that iteration,
and $\rho^{\rm red}_{s,r}(m)$ is the reduced density
matrix defined as
\begin{equation}
\rho^{\rm red}_{s,r}(m) = \sum_{e}
          \langle s,e;m|\hat{\rho}_0 |r,e;m \rangle .
\label{eqn:reduced-dm-def}
\end{equation}
The restricted sum over $r$ and $s$ in (\ref{eqn:time-evolution-intro})
requires
that at least one of these states is discarded at
iteration $m$. The NRG chain length $N$ implicitly
defines the temperature entering Eq.~(\ref{rho-0}):
$T_N \propto \Lambda^{-N/2}$, where $\Lambda > 1$
is the Wilson discretization parameter.

\begin{figure}[tb]
\centering
\includegraphics[width=0.5\textwidth]{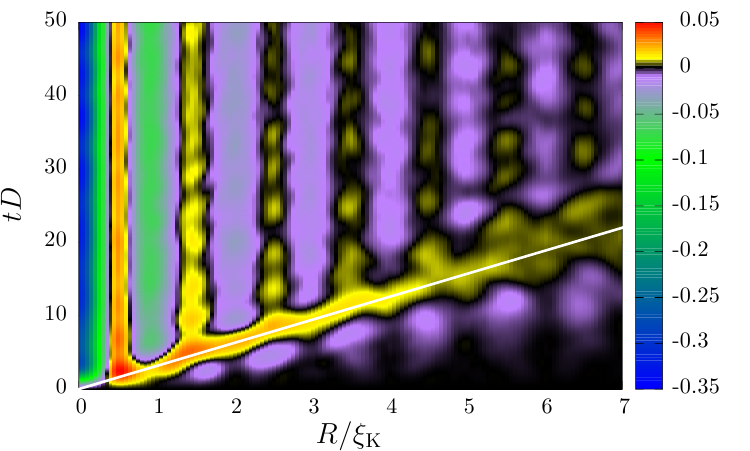}
\caption{(color online)
The 1D time and spatial 
dependent spin-correlation function $\chi(R,t)$ vs 
$x=k_{\rm F} R/\pi$ and $\tau=tD$
for $\rho_0 J=0.3$ 
as a color contour plot. 
Its  color  map is depicted on the right site. 
The correlation propagates with $v_\mathrm{F}$,
which is added as white line as guide to the eye.
NRG parameters are $\Lambda=3$, $N_s=1400$ and $N_z=4$.
}
\label{TDCor_J0.6}
\end{figure}

The derivation of Eq.~(\ref{eqn:time-evolution-intro}) is based on the
two observations. (i) It has been
shown \cite{AndersSchiller2005,AndersSchiller2006} that the set of all
discarded states in the NRG procedure forms a complete basis set of
many-body Fock space which are approximate eigenstates of the
Hamiltonian. (ii) The general local operator $\hat{O}$ is diagonal in
the environment degrees of freedom (DOF) such that the trace over
these DOF can be performed analytically yielding $\rho^{\rm
red}_{s,r}(m)$.  This approach has been extended to steady state
currents at finite
bias \cite{AndersSSnrg2008,SchmittAnders2009,*SchmittAnders2011,JovchevAnders2013}.
The only error of the method stems from the representation of the bath
continuum $H_0$ by a finite-size Wilson chain\cite{Wilson75} and are
essentially well
understood \cite{EidelsteinGuettgeSchillerAnders2012,GuettgeAndersSchiller2013}.

\subsection{Time-dependent spatial correlation function in the Kondo regime}

\label{sec:TD-NRG-Kondo-regime}

After discussing the equilibrium correlation function in Sec.\
\ref{sec:equilibrium}, we present our results for the full
time-depended correlation function $\chi(\vec{r},t)$.  The NRG fixed
point differs for different signs of the Kondo coupling: for $J>0$ the
SC fixed point \cite{Wilson75} is reached while for
$J<0$ the system approaches the LM fixed
point. Therefore, we present data for both regimes and begin with the
antiferromagnetic Kondo coupling.

\begin{figure}[t]

\includegraphics[width=0.5\textwidth]{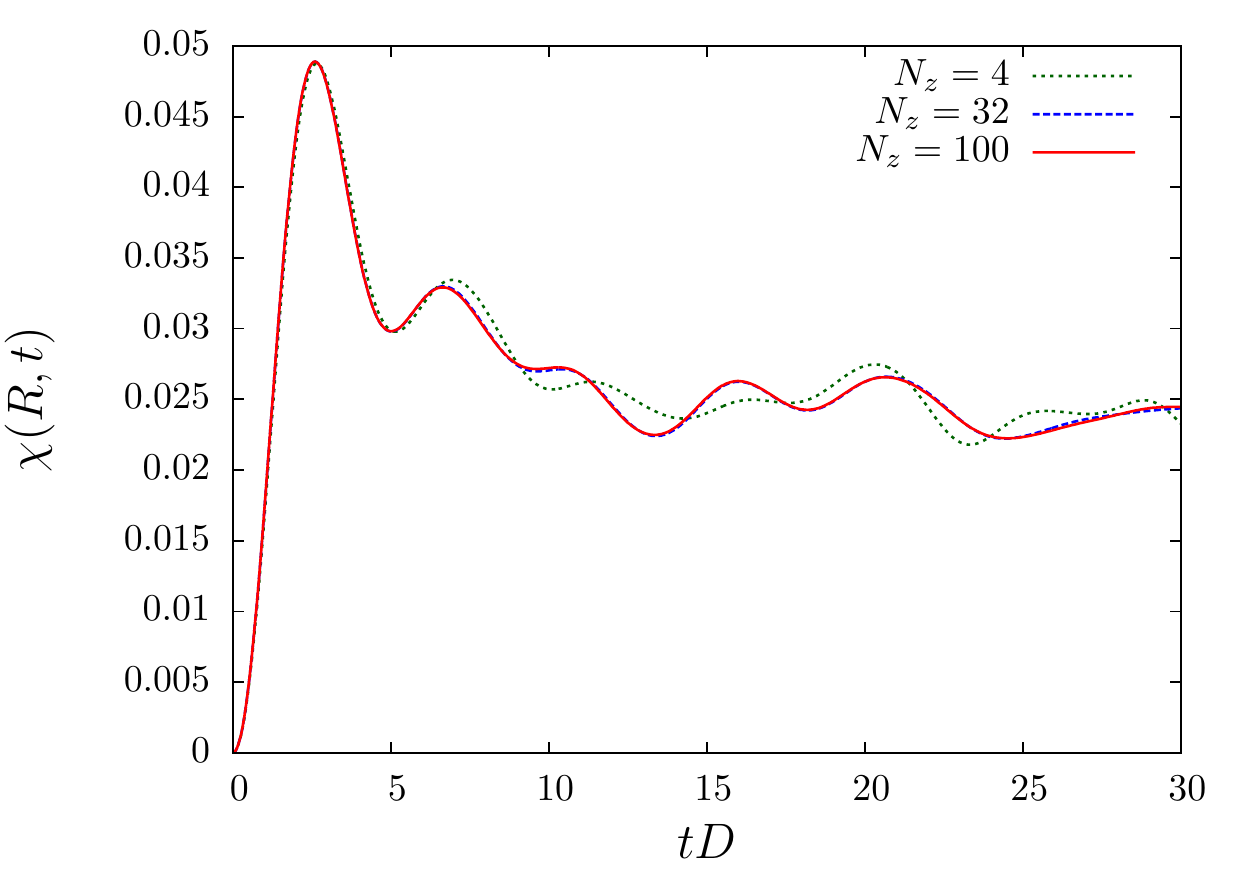}

\caption{(color online)
Time dependent correlation function $\chi(R,t)$ for a fixed distance
$k_F R/\pi=0.51$ vs time for different number of $z$ averages $N_z$ and a
fixed $\rho_0 J=0.3$. 
NRG parameters are $\Lambda=3$ and $N_s=1600$.
}
\label{fig:chi-r-fixed-vs-Nz}
\end{figure}

$\chi(R,t)$ is depicted as a function of the dimensionless
distance $x = k_{\rm F} R/\pi$ and the dimensionless time $\tau
=t D$ for a moderate Kondo coupling $\rho_0 J=0.3$ as a color contour
plot for 1D in Fig.~\ref{TDCor_J0.6}.  Each distance $R$ requires a single
TD-NRG run.  Therefore, we have restricted ourselves to $N_z=4$
values for the $z$ averaging \cite{YoshidaWithakerOliveira1990,AndersSchiller2005,AndersSchiller2006} 
for the $N_R=350$ different values of $R$
included in Fig.\ \ref{TDCor_J0.6} and only kept a moderate number of
NRG states.

The development of the ferromagnetic correlation maximum at $x=
1/2$ is clearly visible already after very short times corresponding
to $\chi_\infty(R)$ depicted in Fig.\ \ref{fig:2ab}(a).  For
$t\to\infty$, the equilibrium Friedel oscillations as discussed above
are recovered: the correlation function has minima  for $x = n$,
and maxima are found  for odd multiple $x=n+1/2$.
For larger distances and times the ferromagnetic correlations are
suppressed in favor of purely antiferromagnetic correlations 
as expected from the equilibrium correlation function.

After the ferromagnetic correlation maximum has passed,
$\chi(R,t)$ exhibits  some weak oscillations in time for $R=$ const. 
In order to discriminate between  finite size oscillations caused by the bath discretiation \cite{AndersSchiller2006,EidelsteinGuettgeSchillerAnders2012,GuettgeAndersSchiller2013}
and  the real nonequilibrium dynamics of the continuum, 
the time evolution of $\chi(R=0.51\pi/k_{\rm F},t)$  is
shown for different number of z-averages $N_z$ in Fig.\ \ref{fig:chi-r-fixed-vs-Nz}. 
Since the short-time oscillations clearly 
converge with increasing $N_z$,  they contain relevant real-time dynamics which will
be analysed in more detail in the next section below.
In the long-time limit, the TD-NRG oscillates around
a time average, which is independent of $N_z$,
and is close to the equilibrium correlation function
$\chi_\infty(R)$. Those oscillations are partially related to  the bath
discretisation \cite{AndersSchiller2006,EidelsteinGuettgeSchillerAnders2012,GuettgeAndersSchiller2013}
and are be suppressed for increasing number of $z$ values $N_z$, decreasing $\Lambda$ and
increasing number of $N_s$.  
Therefore, we conclude that thermodynamic
equilibrium is reached up to the well understood
small discretisation errors 
\cite{AndersSchiller2006,EidelsteinGuettgeSchillerAnders2012,GuettgeAndersSchiller2013}.

\begin{figure}[tb]

\flushleft{(a)}\\
\includegraphics[width=0.5\textwidth]{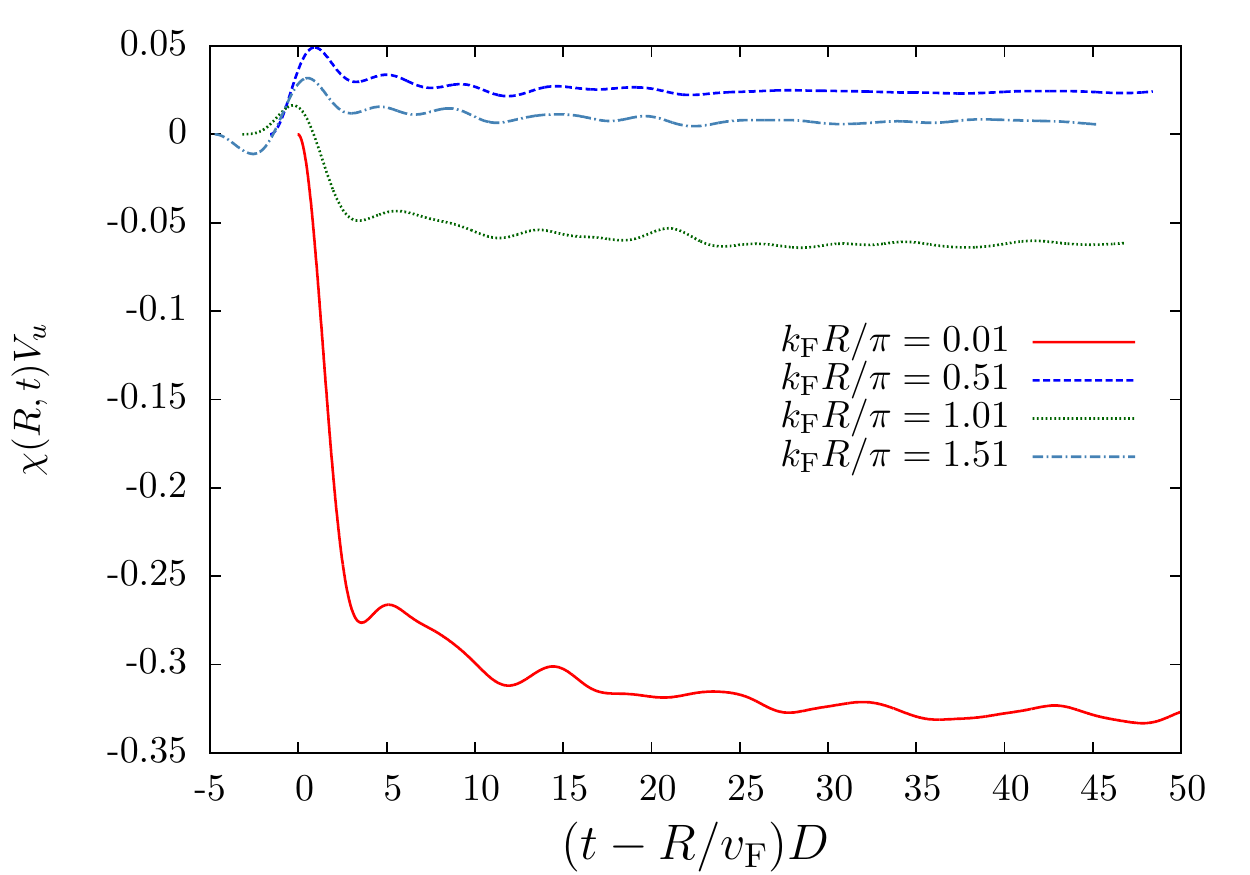}
\flushleft{(b)}\\
\includegraphics[width=0.5\textwidth]{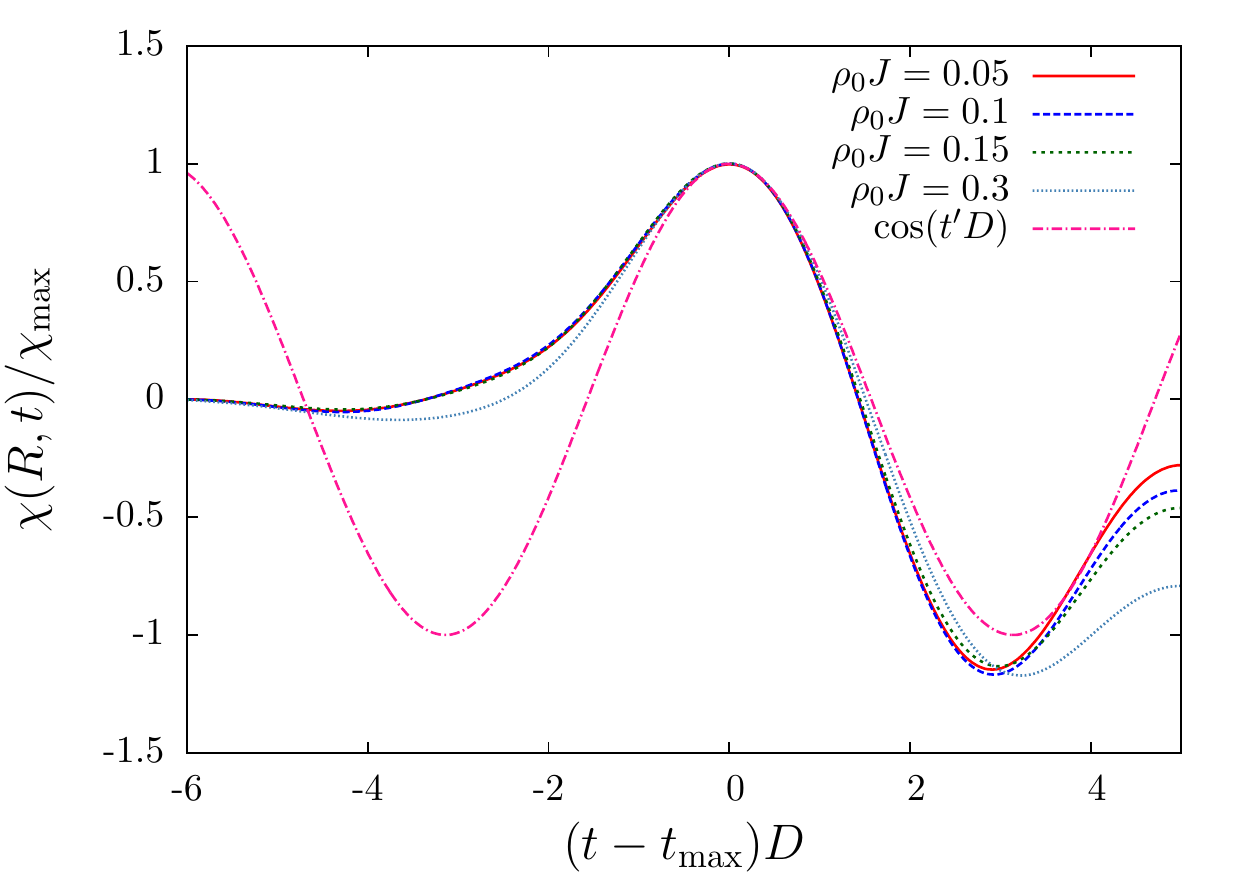}
\caption{(color online)
(a) Time-dependent correlation function $\chi(R,t)$  vs $t'=t-R/v_{\rm F}$
for four different distances $\kr=0, 0.51,1.01.,1.51$
for $\rho_0 J=0.3$ in 1D. 
(b) Rescaled time-dependent correlation function 
$\chi(R,t)/\chi_\mathrm{max}$ vs $t-t_\mathrm{max}$ for $k_{\rm F} R/\pi=2.01$ 
and four different couplings $\rho_0 J=0.05, 0.1 0.15, 0.3$. 
$t_\mathrm{max}$ is the position and $\chi_\mathrm{max}$ is the amplitude of the ferromagnetic peak.
NRG parameters are $\Lambda=3, N_s=1200$ and $N_z=32$.
}
\label{fig:chi-r-t-fixed-R}
\end{figure}

\subsubsection{How are the Kondo correlation building up at different distances $R$
with time?}

Clearly visible is the propagation of ferromagnetic correlations away
from the impurity with a constant velocity given by the Fermi velocity
of the metallic host.  At the impurity site, an antiferromagnetic
spin-spin correlation develops rather rapidly.  Since the total spin
is conserved in the system, a ferromagnetic correlation-wave propagates
spherically away from the impurity spin.  The added white line
$R=v_{\rm F} t$ serves as guide to the eye in Fig.~\ref{TDCor_J0.6} to
illustrate this point.  This line represents the analog to a
light cone in electrodynamics.

Inside the light cone, the equilibrium correlation function is reached
rather fast.  To exemplify this, we plot $\chi(R,t)$ as function of
relative time $t'=t-R/v_{\rm F}$ for four different distances $R$ in
Fig.\ \ref{fig:chi-r-t-fixed-R}(a). Negative $t'$ corresponds to the
correlations outside the light cone, while for $t'>0$ the spin correlation
function $\chi(R,t)$ inside the light cone is depicted.

At the origin of the impurity ($R=0$), a antiferromagnetic
correlation develops \footnote{For a true two-channel calculation, using
the same NRG parameters requires a finite $R$, otherwise $N_o(\e)=0$
and the numerics breaks down.}
on the time scale $1/\sqrt{J}$: the short time dynamics is linear
in the Kondo coupling and proportional to $t^2$ as will be discussed in 
greater detail below.

At $t'=0$ and finite distance $R>0$, a significant ferromagnetic 
correlation-wave  peak is observe which decays rather rapidly. 
Its position  corresponds to the yellow (color-online) 
light cone shown in Fig.\ \ref{TDCor_J0.6}.  
In order to shed some light into the nature of this rapid decay, we
present  the ratio
$\chi(R,t)/\chi_\mathrm{max}$ versus $(t-t_\mathrm{max}) D$ for a
constant distance $\kr/\pi=2.01$ and different couplings $\rho_0 J$
in Fig.\ \ref{fig:chi-r-t-fixed-R}(b).
$\chi_\mathrm{max}$ has been defined as $\chi_\mathrm{max}=\chi(R,t_\mathrm{max})$,
and $t_\mathrm{max}$  is the numerical position of the ferromagnetic peak.
Note that  $t_\mathrm{max} $ slightly
differs from  bare light cone time scale $R/v_\mathrm{F}$ and  is shifted to larger times
with increasing $\rho_0 J$ (not shown.)  This increasing shift can be analytically understood, and
we will give the detailed explanation in Sec.\ \ref{sec:perturbation-theory}
below. 
After dividing out the amplitude $\chi_\mathrm{max}$ of the maximum,
the decay surprisingly  shows a universal behavior and
the time scale is simply given by $1/D$.  
A comparision of the oscillation with
$\cos( t' D)$ (pink dash-dotted line) shows a remarkable agreement
for small times $0<t'  D<1$. 
This indicates that the functional form of $\chi(R,t)$ for fixed $R$ 
consists of a damped oscillatory $\cos( t'  D)$ term 
whose maximum is reach when the ferromagnetic correlation wave reaches
the distances $R$ at the time  $t_{\rm max}> R/v_{\rm F}$. 
For larger times  $t'$, $\chi(R=2.01\pi/K_{\rm F})$ has to approach 
an finite antiferromagnetic value. Therefore, the oscillations in the TD-NRG
are not centered  around the origin but shifted to negative values 
as can seen in Fig.\ \ref{fig:chi-r-t-fixed-R}(b) by comparing with 
the undamped $\cos( t'  D)$ curve.

Most striking, however, is the building up of correlations for $t'<0$
outside of the light cone. These correlations are 
antiferromagnetic and show no exponential decay.  These correlations
appear shortly before the light cone. They reach their largest modulus
for odd multiple $k_\mathrm{F}r/\pi = n+1/2$ and decay with a power
law as $t D$ goes to zero. In Sec.\ \ref{Sec_intrinsic} below, we
will provide a detailed analysis of their origin and present an
analytically calculation in $J$ that agrees remarkably well with the
observed TD-NRG results.

Such a building up of correlations outside of the light cone has recently
be reported in a perturbative calculations \cite{Medvedyeva2013}
neglecting, however, the $2k_{\rm F}$ oscillations.  Here, we present results for
a full nonperturbative calculation which includes the Friedel
oscillations containing the RKKY mediated effective spin-spin
interaction.

\begin{figure}[tb]
  \flushleft{(a)} \center{$\rho_0J=0.15$}
  \includegraphics[width=0.5\textwidth]{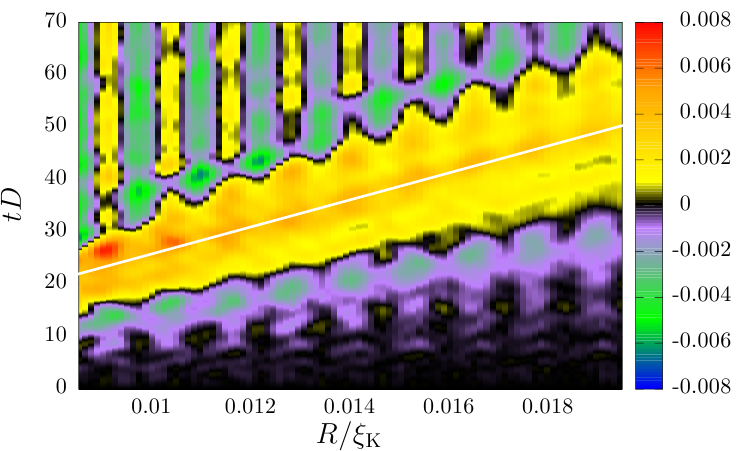}
  \flushleft{(b)} \center{$\rho_0J=0.3$}
  \includegraphics[width=0.5\textwidth]{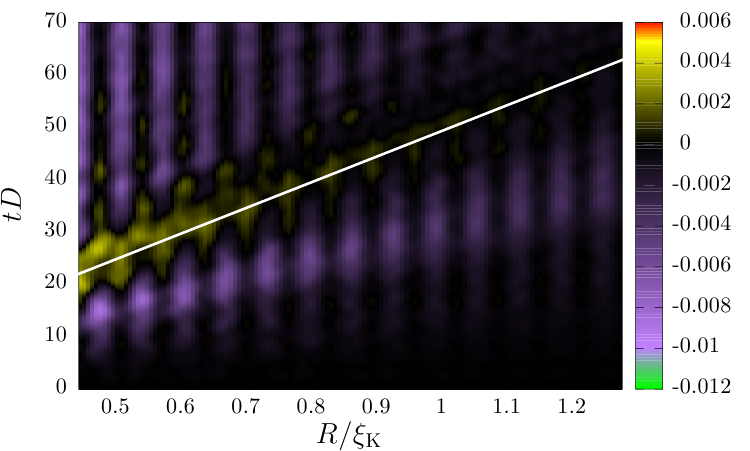}
  
  \caption{(Color online)
  The time-dependent correlation $\langle \vec{S}_{\rm imp}\vec{s}(R)\rangle(t)$ for different distances as a color contour plot. 
Its  color  map is depicted on the right site. 
  (a) The oscillation between ferromagnetic and antiferromagnatic correlations for long times is only observed for short distances $R/\xi_\mathrm{K}\ll1$. 
  (b) For long times, oscillations only between zero and antiferromagnetic correlations are observed. 
The ferromagnetic propagation vanishes at around $R/\xi_\mathrm{K}\approx 1$. 
Both long-time behaviors are in agreement with the NRG equilibrium results.
NRG parameters are $\Lambda=3, N_s=1400$ and $N_z=32$.
  }

  \label{TD_xi_Kl}
\end{figure}

The different behavior for short and long distances is  illustrated
in Fig.\ \ref{TD_xi_Kl}. The upper panel shows results for short distances  $r/\xi_\mathrm{K} \ll 1$. 
We observe the distinctive ferromagnetic correlation which propagates with the 
Fermi velocity through the conduction band. 
Inside the light cone, we find oscillations between ferromagnetic and antiferromagnatic correlations. 
In the lower panel, the behavior for longer distances is depicted. 
We find that the ferromagnetic propagation vanishes at around $\kr\approx \xi_\mathrm{K}$. 
At these distances, we only observe oscillations between zero and antiferromagnatic correlations inside the light cone. 
For both cases, the long-time behavior agrees remarkable well with the NRG equilibrium calculations.

\subsection{Perturbation theory}
\label{sec:perturbation-theory}

Surprisingly, we found in our TD-NRG results  the building up of spin-correlations outside of the light cone which do not decay
exponentially. In order to rule out TD-NRG artifacts and shed some
light into its  origin, we perturbatively calculate $\langle \vec{S}_{\rm imp}
\vec{s}(r) \rangle(t)$ up to second order in $J$.
 
Since only $H_0$ enters the initial density operator, we transform
all operators into the interaction picture and, after integrating the von
Neumann equation, we obtain  for the density operator
\begin{eqnarray}
\label{eq:von-neumann} \rho^I(t)&\approx & \rho_0+i \int_0^t \left[
\rho_0,H_{K}^{I}(t_1) \right] dt_1 \nonumber \\ &&-
\int_0^t\int_0^{t_1}\left[\left[\rho_0,H_{K}^{I}(t_2)\right]
H_{K}^{I}(t_1)\right] dt_2 dt_1
\end{eqnarray}
which is exact  in second order in the Kondo coupling $J$. The
superscript $I$ labels the operators in interaction picture
$A^I(t)=e^{iH_0t}Ae^{-iH_0t}$, and the boundary condition
is given by $\rho_0=\rho^I(t=0)$. We use
this $\rho^I(t)$ to calculate the spin-spin correlation function
\begin{eqnarray} \chi(\vec{r},t) &=& {\rm
Tr}\left[\rho^I(t)\vec{S}_{\rm imp} \vec{s}^{I}(\vec{r},t) \right]
\end{eqnarray} 
where only expectation values with respect to the initial density operator $\rho_0$
enter. The
occurring commutators are cumbersome but can be evaluated
analytically (for details see Appendix \ref{sec:perturbation-Simp-s-bad}).
We note that  the contribution in linear order in $J$ does not vanish
for a perturbation  $H_K= J\vec{S}_{\rm imp}\vec{s}(0)$. 
Although the time
integrals can be performed analytically, the multiple momenta
integrations of the free conduction electron states must be performed
numerically.

\begin{figure}[t]
\centering
\includegraphics[width=0.5\textwidth]{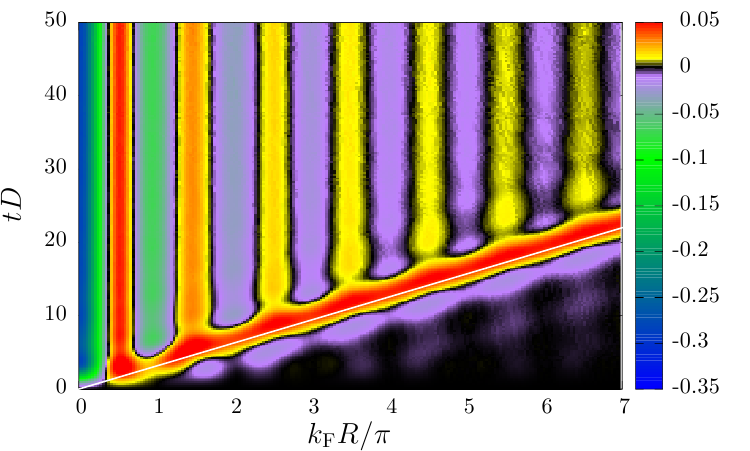}
\caption{(Color online) 
The analytical $\chi(R,t)$ evaluated numerically   for $\rho_0 J=0.3$
up to second order in $J$
as a color contour plot. 
Its  color  map is depicted on the right site. 
The light cone $R=v_\mathrm{F} t$ has been
added as a white line.
}
\label{FirstAndSec_Order}
\end{figure}

The sum of the first- and second-order contributions to $\chi(R,t)$ are
shown in Fig.\ \ref{FirstAndSec_Order} up to relatively large times.
This perturbatively calculated $\chi(R,t)$ turns out to be well behaved and does not
contain secular terms. Clearly, the Kondo physics is not included in
such an approach remaining only valid for
$r/\xi_\mathrm{K}\ll1$ and $t T_K \ll 1$.  Therefore, we expect
deviations at large distances and times from the NRG results.

Nevertheless, the NRG result depicted in Fig.\ \ref{TDCor_J0.6} and the
perturbation theory result  qualitatively agrees very well.  As in the
TD-NRG, ferromagnetic correlations propagate spherically away from the
impurity with Fermi velocity (white line as guide to the eye) in the
perturbative solution.  For long times an equilibrium is reached:
we recover the distance dependent Friedel oscillations which are known
from the RKKY interaction.  We also find the same antiferromagnatic
correlations outside the light cone as in the NRG results. Again,
their maxima are located at odd multiple of $x=\kr/\pi= n +1/2$ at the
same positions as predicted in the TD-NRG calculations.

To the leading order in the Kondo coupling $\rho_0 J$, 
the ferromagnetic  wave in the correlation
function propagates on the light cone line $R=v_{\rm F} t$.
Note, however,  that the  peak position of the analytical $\chi(R,t)$ 
plotted  in  Fig.\ \ref{FirstAndSec_Order} 
is slightly shifted to  later times than defined by  the light cone line.
This shift  coincides with the one observed in the TD-NRG results shown in
the Figs.\ \ref{TDCor_J0.6} and  \ref{fig:chi-r-t-fixed-R}. 
With our analytical analysis at hand, we can provide a detailed understanding
of this effect.
The first-order contribution given by Eq.\ (\ref{FirstOrder})   
yields a peak of ferromagnetic correlations positioned  exactly on  light cone line.
However, the maximum of the second-order contribution (\ref{SecOrder}) is shifted to slightly larger 
times.  Adding both contributions generates  a $J$-dependent line for the ferromagnetic peak position
away from the light cone line: 
the larger $J$, the later the ferromagnetic maximum occurs  
due to increasing importance of
the second order contribution.

While these spatial oscillations are implicitly encoded in the
effective even and odd density of states in the NRG calculation, they
are explicitly generated by the momenta integration in the
perturbative approach. This confirms our TD-NRG results and provides a
better understanding of the numerical data.

Comparing Figs.\ \ref{TDCor_J0.6}  and \ref{FirstAndSec_Order} in more
detail illustrates the shortcomings of the perturbative approach which
remains only valid for $R/ \xi_\mathrm{K}\ll 1$. As discussed in 
Sec.\ \ref{sec:equilibrium} above , the decay of the envelope function
crosses over from a  $1/R$ to an $1/R^2$ behavior due to the
Kondo screening of the local moment for $R/\xi_{\rm K}>1$. 
Since the  perturbative approach is unable to access the Kondo-singlet formation
the perturbative solution depicted in Fig.\ \ref{FirstAndSec_Order}
decays as $1/R$ at all distances and also remains oscillating
between ferromagnetic and antiferromagnetic correlations while
the TD-NRG correctly predicts only antiferromagnetic correlations
inside the light cone once $R$ exceeds the Kondo length scale $\xi_{\rm K}$.

\begin{figure}[tb]
  \includegraphics[width=0.5\textwidth]{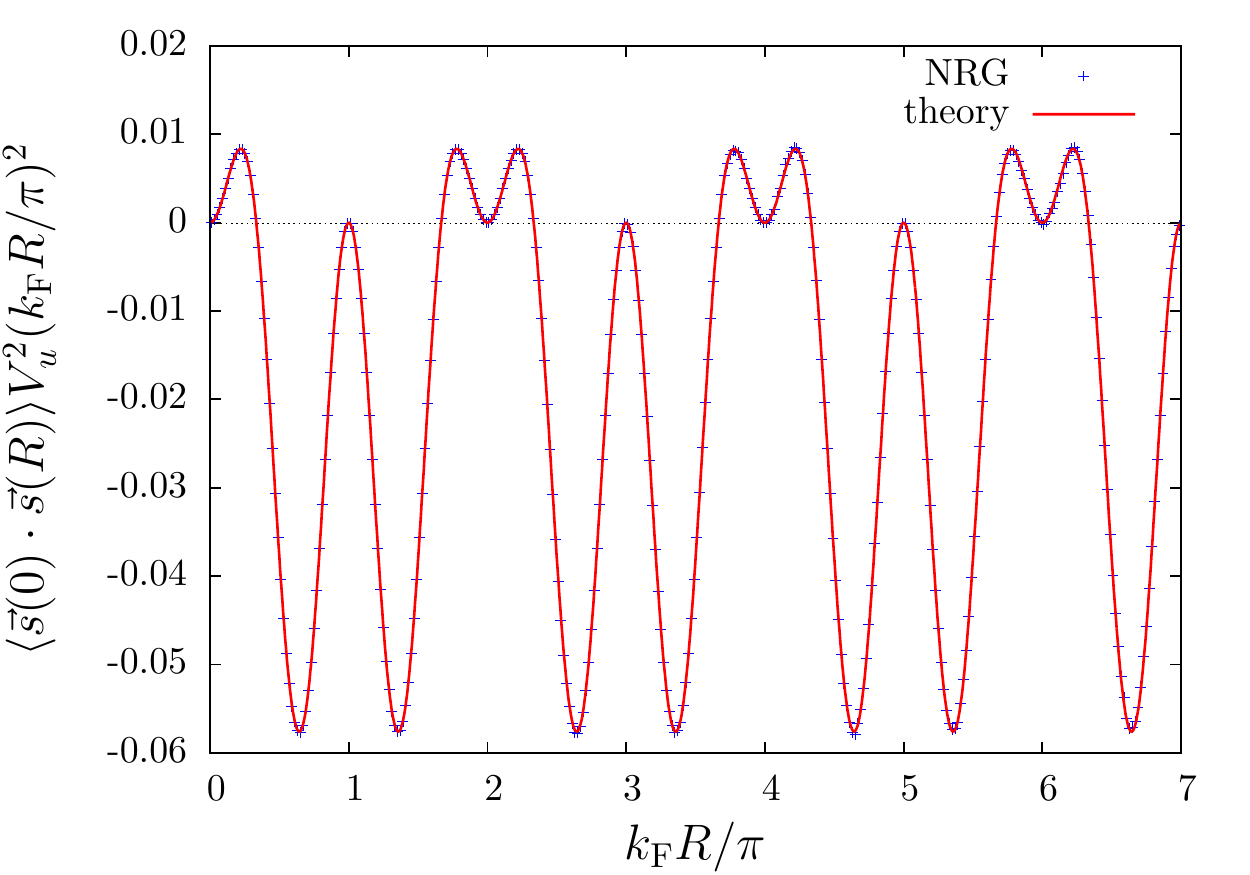}
  \caption{(Color online) The intrinsic spin-spin correlations of
the Fermi sea between the spin densities $\vec{s}(0)$ and $\vec{s(r)}$
in 1D.  Via the mapping to the even and odd conduction bands we are
able to measure bath properties at distance $R$ and get a perfect
agreement between theory and NRG results.
NRG parameters are $\Lambda=3$ and $N_s=1200$.  
}
  \label{compare_intr_Taylor}
\end{figure}

\subsection{Intrinsic correlations of the Fermi sea}\label{Sec_intrinsic}

Since the perturbative results agree remarkably well with the TD-NRG
data for short distances, the analytical approach can be used
to gain an explicitly understanding  of the 
correlations outside of the light cone. 
It has been  proposed \cite{Medvedyeva2013} that these correlations 
originate from the intrinsic spin-spin correlations in the Fermi sea 
$\langle \vec{s}(0)\vec{s}(\vec{r}) \rangle$ already present prior to 
the coupling of the impurity. Once the impurity is coupled to the local 
conduction electron  spin density at time $t=0$, 
we instantaneously  probe these intrinsic entanglements of the Fermi sea
between the local spin density and the spin density at a large distance $R$.

For $J=0$,  $\langle \vec{s}(0)\vec{s}(R) \rangle$ can be calculated 
analytically and is given by
\begin{eqnarray}
	\langle \vec{s}(0)\vec{s}(R) \rangle& =&
	\frac{3\sin(k_\mathrm{F}R)\sin(\frac{k_\mathrm{F}R}{2})\cos(\frac{3}{2}k_\mathrm{F}R)}{4V_u^2(k_\mathrm{F}R)^2}
\end{eqnarray}
in 1D. This exact result coincides with the NRG data obtained by
setting $J=0$ in an equilibrium NRG calculation as shown in Fig.\
\ref{compare_intr_Taylor}. This excellent agreement between the
analytical and the NRG approaches serves as further evidence for the
numerical accuracy of mapping Eqs.\ (\ref{eqn:7})-(\ref{eqn:9}) to the
two discretized and properly normalized Wilson chains for even and odd
parity conduction bands.

\begin{figure}[tb]
  \includegraphics[width=0.5\textwidth]{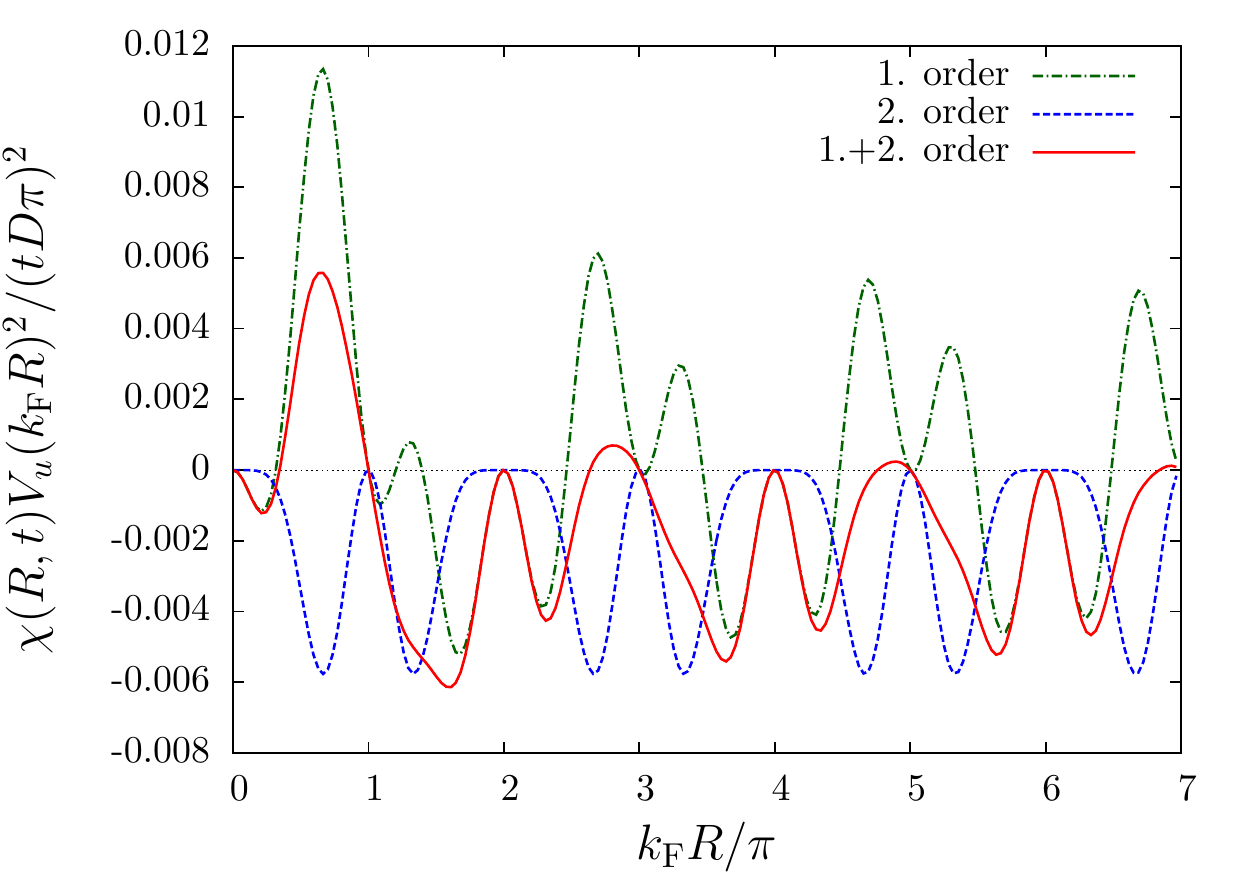}
  \caption{(Color online) 
First and second order
of $\chi(R,t)$ 
for small times $t$ and
$\rho_0 J=0.3$ in 1D.  Even for short times we observe correlations for
large distances. A comparison with a) shows that the positions of
these correlations coincide with the positions of the intrinsic
correlations of the Fermi sea.  So the correlations outside the light
cone originate from these intrinsic correlations.  
 }
  \label{fig:10}
\end{figure}

In order to connect the intrinsic spin entanglement of the decoupled
Fermi sea with the observed antiferromagnetic correlations outside of
the light cone, we expand the perturbative calculated $\chi(R,t)$ for
small times $0\le t D\ll 1$ and perform the momentum integrations
analytically, separately for the first and the second order contributions.

The leading order
terms in time are proportional to $\propto (t D)^2$ and decay as
$1/R^2$. We recall the difference between the $1/R^2$ 
decay outside of the light cone for the short-times dynamics
and the well understood $1/R$ decay inside 
the light cone  when the equilibrium reached.
Therefore, we have plotted the perturbative results in 1D 
as $\chi(R,t) (R/ (t D))^2$  in Fig.\ \ref{fig:10} 
to eliminate the time dependence and compensate for the leading 
spatial decay of the envelop function.  
The  first-order contribution (green online) is shown as a dashed-dotted line,
the second-order contribution  (blue online) is depicted as a dashed line,
and the sum of both (red online) is added as a solid line for $\rho_0J=0.3$  
in Fig.\ \ref{fig:10}.

Since the second-order contribution to 
$\chi(0,t)$ remains always zero in 
a short-time expansion, the time evolution of the
antiferromagnetic correlation at $R=0$ is dominated by the first-order 
term being proportional to $\propto J t^2$.  Therefore, the time scale for the initial fast
buildup of the local antiferromagnetic correlation  is
given by $1/\sqrt{J}$ confirming 
the TD-NRG short-time dynamics  for $k_{\rm F} R/\pi=0.01$ as depicted in
Fig.\ \ref{fig:chi-r-t-fixed-R}(a).

The largest contribution for short times stems from the ferromagnetic
peak around $\kr/\pi=0.5$. However, correlations are visible at all
length scales which develop quadratically in time.  The positions of
the maxima and minima agree remarkable with those of the intrinsic
correlation function $\langle \vec{s}(0)\vec{s}(R) \rangle$ of the
Fermi sea depicted in Fig.\ \ref{compare_intr_Taylor} and both decay
with $1/R^2$.

Since the first-order contribution is sensitive to the sign of $J$, its
maxima contribute with the equal sign as $\langle \vec{s}(0)
\vec{s}(R) \rangle$ for ferromagnetic $J$ and opposite sign for an
antiferromagnetic coupling.
The second-order term only adds negative (antiferromagnetic)
contributions to $\chi(R,t)$. The location of its
negative peaks coincide with the antiferromagnetic peak location of
the spin-spin correlation function $\langle \vec{s}(0)\vec{s}(\vec{r})
\rangle$ of the decoupled Fermi sea depicted in Fig.\
\ref{compare_intr_Taylor}.
The sum of both orders contains only small ferromagnetic correlations
for larger distances $\kr/\pi >1$ for  $\rho_0 J=0.3$.
The larger the coupling $J$ is the smaller these ferromagnetic
correlations are due to the increasing dominance of the second-order
term.

\begin{figure}[t]
\centering

  \flushleft{(a)}
\includegraphics[width=0.5\textwidth]{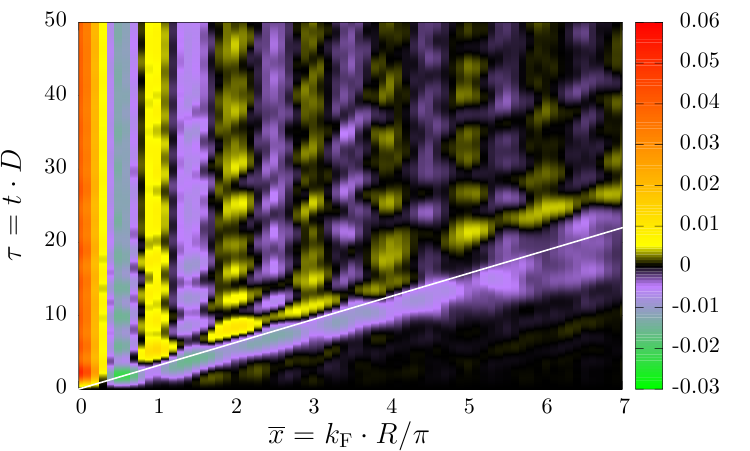}

  \flushleft{(b)}
\includegraphics[width=0.5\textwidth]{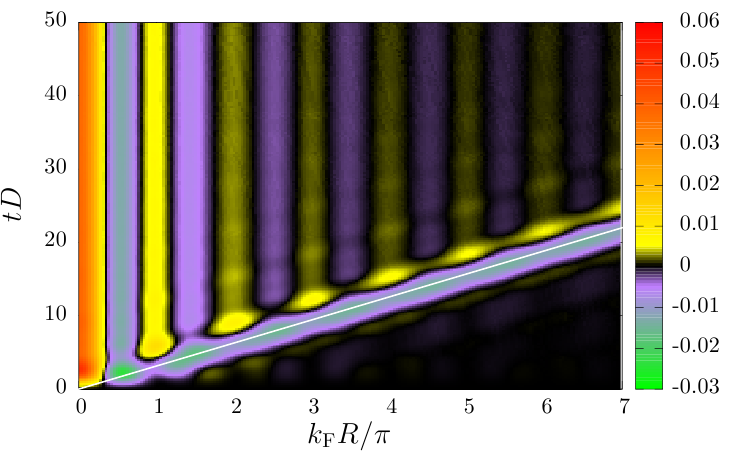}

\caption{(Color online)
(a) The time-dependent spin correlation function $\chi(R,t)$ vs $x=k_{\rm F} R/\pi$ and $\tau=tD$
for a ferromagnetic Kondo coupling $\rho_0 J=-0.1$ in 1D
as color contour plot. 
Its  color  map is depicted on the right site. 
(b)
The analytic spin correlation function $\chi(R,t)$ for the same parameter
calculated in second-order perturbation theory in $\rho_0 J$.
NRG parameters are $\Lambda=3$, $N_s=1400$ and $N_z=32$.
}
\label{fig:chi-ferro-j03}
\end{figure}

We can conclude from this detailed analytical analysis that the
antiferromagnetic correlation directly in front of the light cone
results from the antiferromagnetic peak in the intrinsic correlations
from the Fermi sea at around $\kr\approx 1.6\pi$.  This peak also
propagates through the conduction band with the Fermi velocity.
Because the intrinsic correlations decay with $1/(\kr)^2$, the
propagation of the antiferromagnetic peaks for larger distances 
is not visible.

\subsection{Local moment regime: ferromagnetic coupling: $J<0$}
\label{sec:LM-TD-NRG}

Now we extend the discussion to ferromagnetic Kondo couplings. In this
regime, the LM fixed point is stable, and the ground state
is two fold degenerate in the absence of an external magnetic
field. In the RG process, the Kondo coupling is renormalized to
zero. Nevertheless, the spatial spin-correlation function
$\chi_\infty(R)$ remains finite for $T\to 0$ as discussed  in Sec.\
\ref{sec:ferro-J-equ}.

The results for the time-dependent spin-correlation function $\chi(R,t)$
are shown as a color contour plot in Fig.\
\ref{fig:chi-ferro-j03}. Figure \ref{fig:chi-ferro-j03}(a) depicts the TD-NRG calculation,
while the analytical result obtained up to the second-order perturbation
theory is added as panel (b) for the same parameters.  Since the 
first-order contribution is sign sensitive, the analytical correlation function
differs significantly from the antiferromagnetic regime displayed in
Fig.\ \ref{FirstAndSec_Order}.

As in the Kondo regime, the analytical and the TD-NRG results agree
qualitatively very well. The Friedel oscillations with the frequency
$2k_{\rm F}$ are clearly visible inside the light cone. Note the phase
shift compared to the Kondo regime: now the ferromagnetic correlations
are observed at $x= n$ and the antiferromagnetic correlations at
half integer values of $x$.
 
Since a ferromagnetic correlation is building up at the impurity spin
position on a very short time scale $\propto 1/\sqrt{J}$, and the
total spin of the system is conserved, an antiferromagnetic
correlation wave spherically propagates away from the origin traveling
with the Fermi-velocity $v_{\rm F}$ (again we have added a white line
$R=v_{\rm F} t$ as guide to the eye to both panels). 
For ferromagnetic couplings the peak position of the propagation is slightly shifted to earlier times $t_\mathrm{max}<R/v_\mathrm{F}$,
due to the sign change of the first-order contribution.
The correlations
outside the light cone are stronger suppressed compared to the
Kondo-regime. Again, we can trace the origin to the intrinsic
entanglement of the Fermi sea by consulting Figs.\
\ref{compare_intr_Taylor}  and \ref{fig:10} as well as the discussion above.

\subsection{Finite temperature: cutoff of the Kondo correlations}
\label{sec:finite-temp}

\begin{figure}[t]
\centering

  \flushleft{(a)}
 \includegraphics[width=0.5\textwidth]{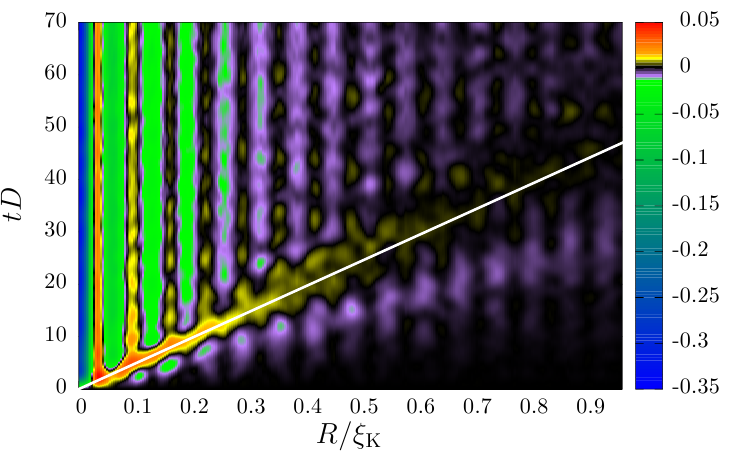}

  \flushleft{(b)}
\includegraphics[width=0.5\textwidth]{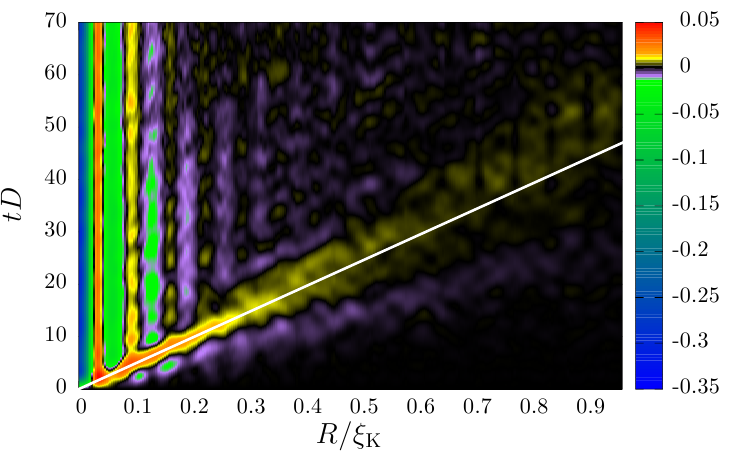}

\caption{(Color online)
The correlation function $\chi(R,t)$ vs $x= R/\xi_\mathrm{K}$ and $\tau=tD$ 
for the medium coupling $\rho_0 J=0.3$ and different temperatures in 1D.
A comparison between the correlation function for (a) zero temperature and
(b) $T=2T_\mathrm{K}$ reveal that correlations in and outside of the light cone
vanish at around $x= R/\xi_\mathrm{K}=0.5$ for $T=2T_\mathrm{K}$. 
NRG parameters are $\Lambda=3$, $N_s=1400$ and $N_z=4$.
}
\label{fig:finitetemp}
\end{figure}

Up to  now, we have only considered the zero-temperature limit.
Next, we extend the discussion to the propagation of the correlations at 
finite temperatures.

Figure \ref{fig:finitetemp}  illustrates the difference in the time-dependent 
correlation function at the two different temperatures $T=0$ and $T=2T_\mathrm{K}$,
depicted in panels (a) and (b), respectively.
Note that we have used the data of Fig.~\ref{TDCor_J0.6} but  measure the distance $R$ in
units of the Kondo correlation length $\xi_\mathrm{K}$. 

For both temperatures, we observe the correlations in and outside of the light cone and the propagation
of the ferromagnetic correlation with Fermi velocity as described in Sec.~\ref{sec:TD-NRG-Kondo-regime}.
While for small distances $R/\xi_\mathrm{K} \leq 0.15$, 
the spin-correlation functions} agree  well for both temperatures,
the correlations inside and outside of the light cone become strongly suppressed in the finite temperature data 
once $R$ exceeds $0.5 \xi_\mathrm{K}$:
the correlations are cut off at the thermal length scale 
$\xi_\mathrm{T}=v_\mathrm{F}/T=v_\mathrm{F}/(2T_\mathrm{K})=0.5\xi_\mathrm{K}$.

The ferromagnetic correlation which propagates with Fermi velocity, however,
is amplified due to spin conservation. Because of the strong suppression outside of the light cone,
the total spin has to be distributed over a larger area.

\begin{figure}[t]
\centering
 \includegraphics[width=0.5\textwidth]{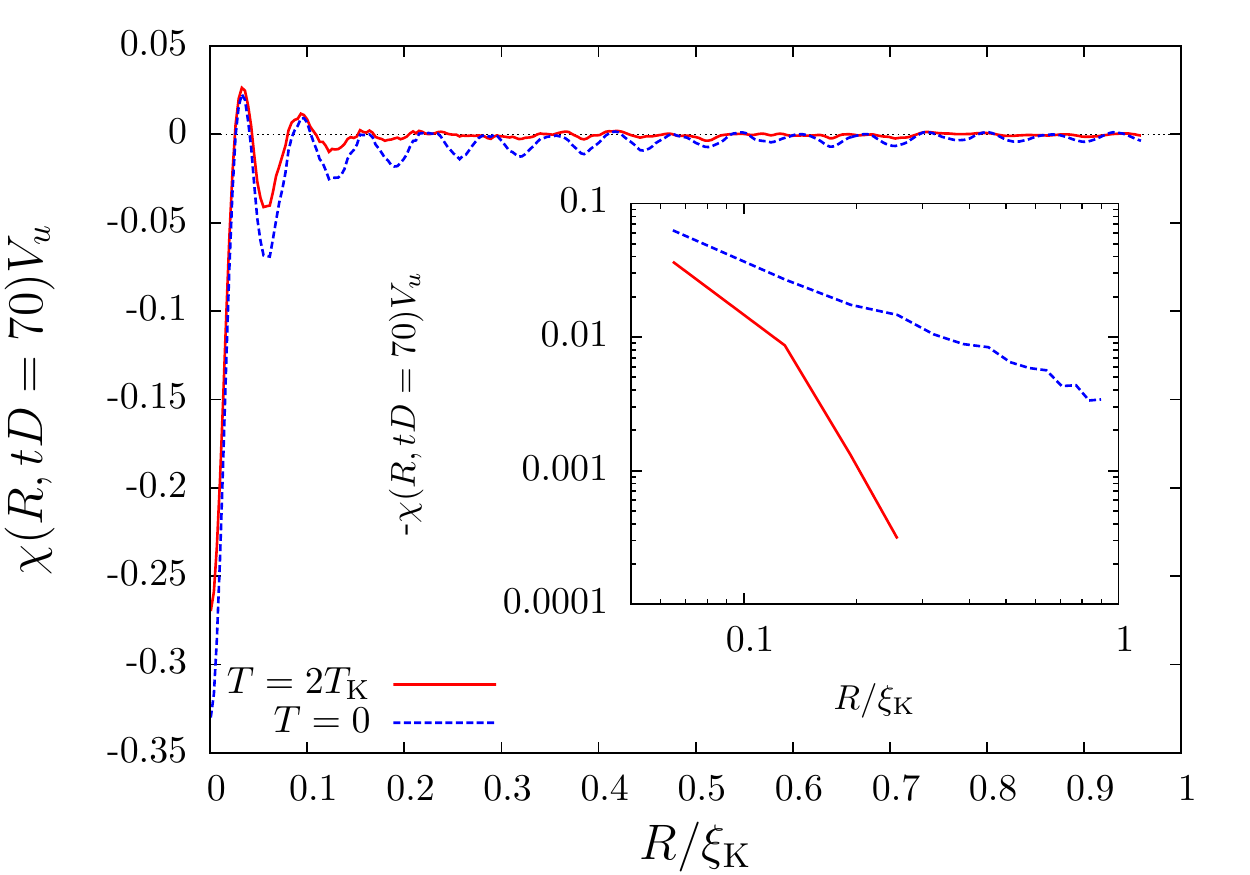}
\caption{
The correlation function $\chi(R,tD=70)$ vs $x= R/\xi_\mathrm{K}$ for the medium coupling $\rho_0 J=0.3$ and the constant time $tD=70$ in 1D.
For $T=0$ the RKKY like oscillations are observed while for $T=2T_\mathrm{K}$ the correlations are cut off.\\
The inset shows the envelope function of $\chi(R,tD=70)$.
}
\label{fig:reach-equi}
\end{figure}

Figure \ref{fig:reach-equi} shows the approach to the equilibrium correlation functions
at large times: the spatial dependence of the spin correlation functions  for both temperatures 
is plotted using the data of Fig.~\ref{fig:finitetemp} at the largest time $D=70$.
In the  $T=0$ curve (dashed line), the RKKY type of oscillations between anti-ferromagnetic  and ferromagnetic 
peaks for small distances and only antiferromagnetic correlations for larger distances are clearly
visible. The decay of its envelope functions plotted as an inset of Fig. \ref{fig:reach-equi} reveals
the power-law decay with the distance.

For the finite temperature $T=2T_\mathrm{K}$, however, the correlations are cut off, and, due to the rapid decay, only a few oscillations (solid red line) can be observed.
The envelope in the inset shows the expected exponential decay at short distances. At larger distances,
the numerical noise of the TD-NRG exceeds the rapid suppression of the correlation function, and one needs to
resort to the equilibrium NRG (see Fig.~3 in  Ref.\ \cite{Borda2007}).

\section{Response function}
\label{sec:response-function}

Within the linear-response  theory,  we can investigate the 
the conduction electron spin-density polarization $\expect{\vec{s}(R,t)}$
as a function of an externally applied local magnetic field $\vec{B}(t)$. Since the Kondo Hamiltonian
is rotational invariant in spin space, the retarded spin-spin susceptibility tensor is diagonal
and proportional to the unity matrix. Therefore, it is sufficient to investigate the 
conduction electron spin-density polarization in $z$ direction $\expect{s^z(R,t)}$ 
at distance $R$ of the Kondo spin 
caused by applying a local magnetic field $\vec{B}(t)= B(t) \vec{e}_z$  
acting on the Kondo spin with $\vec{e}_z$ being the unity vector in
$z$ direction.  

The spin-density polarization $\expect{\vec{s}(R,t)}$ 
\begin{eqnarray}
\label{eqn:27}
\expect{s^z(R,t)} &=& \expect{s^z(R,t=-\infty} \non
&&  +\int_{-\infty}^{\infty} dt'\, \chi^r_{\rm imp-c}(R,t-t') \Delta(t') 
\, ,
\label{eq: convolution}
\end{eqnarray}
is  related to the retarded spin susceptibility 
\begin{eqnarray}
\chi^r_{\rm imp-c}(R,t) &=& -i\expect{
[s^z(R,t), S^z_{\rm imp}]
} 
\Theta(t)
\end{eqnarray}
that is a true response function. Since the system is unpolarized at $t=-\infty$,
$\expect{s^z(R,t=-\infty}=0$ and can be neglected.
In Eq.~\eqref{eqn:27} the local 
applied time-dependent Zeeman splitting  $\Delta(t)=g\mu_B B(t)$ has entered.
Since the spin-density $s^z(R,t)$ 
and $S^z_{\rm imp}$ are  Hermitian operators, $\chi^r(R,t)$
is a purely real function depending only on the spectrum \cite{PetersPruschkeAnders2006} 
$\rho^r_{\rm imp-c}(R,\w) = -\lim_{\delta\to 0^+}\Im  \chi^r_{\rm imp-c}(R,\w +i\delta)/\pi$:
\begin{eqnarray}
\label{eq:sus-imp-bath}
\chi^r_{\rm imp-c}(R,t) &=& -2\int_{0}^{\infty} d\w \rho^r_{\rm imp-c}(R,\w) \sin(\w t)
\, .
\end{eqnarray}

\subsection{Retarded host susceptibility $\chi^r_{\rm c-c}(R,t)$}
\label{sec:Retarded host susceptibility}

The retarded equilibrium host spin-density susceptibility  
\begin{eqnarray}
\chi^r_{\rm c-c}(R,t) &=& -i\expect{
[s^z(R,t), s^z(0,t)]
} 
\Theta(t)
\end{eqnarray}
 can be analytically calculated in the absence of a coupling to the impurity ($J=0$)
 (see Appendix \ref{sec:app-c}). For a 1D linear dispersion, we have obtained 
\begin{eqnarray}
\label{eq:spin-sus-fermi-sea}
\rho^r_{\rm c-c}(R,\w) =&& \frac{1}{2\pi V_u^2N^2} \sum_{k_1,k_2} [f(\epsilon_{k_2})-f(\epsilon_{k_1})]
\nonumber \\
		        && \times \left[ \pi \cos((k_2-k_1)R) \delta(\omega -(\epsilon_{k_1}-\epsilon_{k_2})) \right.\nonumber  \\
			&& + \left. \frac{\sin((k_2-k_1)R)}{\omega -(\epsilon_{k_1}-\epsilon_{k_2})} \right]
\punkt
\end{eqnarray}
The analytic expression \eqref{eq:spin-sus-fermi-sea}
of the spin-spin susceptibility contains the dimensionless frequency  
$k_F R$ for a linear dispersion
causing increasing frequency oscillations with increasing $R$.

\begin{figure}[t]
\centering

  \flushleft{(a)}
   \includegraphics[width=0.5\textwidth]{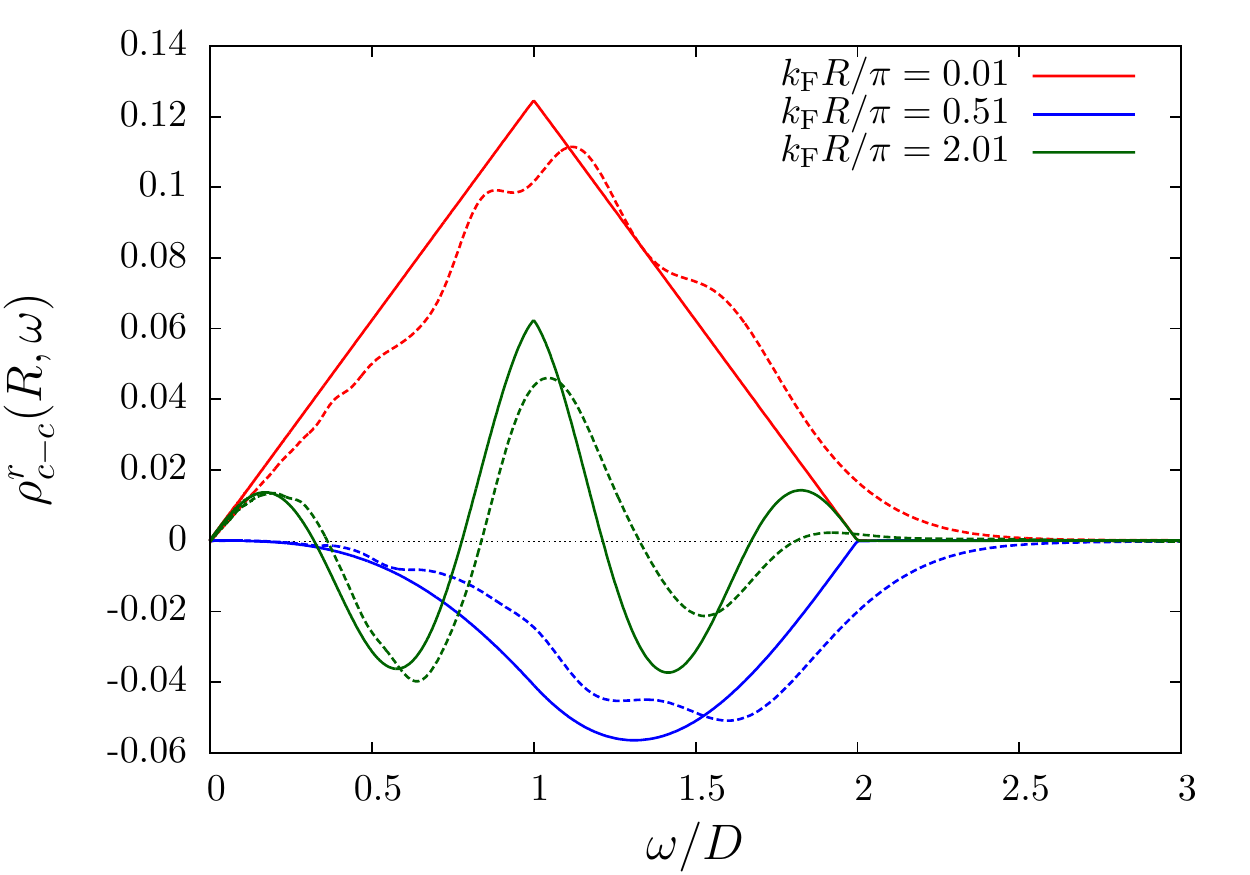}

  \flushleft{(b)}
   \includegraphics[width=0.5\textwidth]{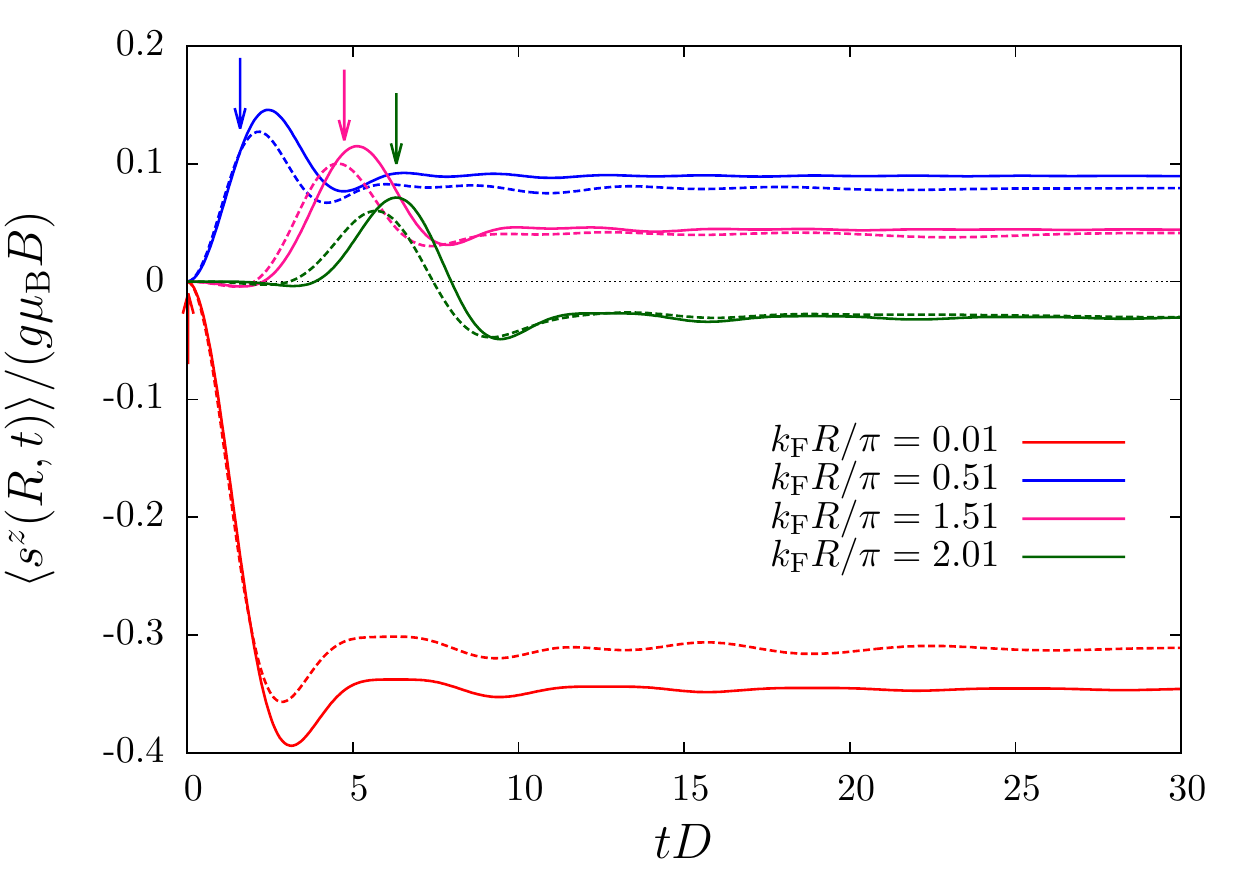}

\caption{(Color online)
(a) The spectral function of retarded host spin-density susceptibility $\chi^r_{\rm c-c}(R,t)$
in the absence of
the impurity for four different distances $k_{\rm F} R/\pi=0.01,0.51,1.51,2.01$. 
The solid line shows the exact analytic result and the dashed line the numerical NRG result.
(b)
The conduction electron spin-density $\expect{s^z(R,t)}$ vs time 
as a response to the local magnetic field $B(t)= B \theta(t)$
applied to the spin-density $s^z(0)$ at the origin. 
The  Fourier transformation
of the spectral information depicted in (a) has been used in combination with 
Eq.~\eqref{eqn:27} where $\chi^r_{\rm imp-c}(R,t)$ has been replaced by $\chi^r_{\rm c-c}(R,t)$. 
The arrows indicate the time $\tau=R/v_{\rm F}$. Note that the $\expect{s^z(R,t)}$ has been normalised
to the Zeeman energy $\Delta_0=g\mu_\mathrm{B}B $.
NRG parameters: $\Lambda=2.25$, $N_s=3000$, $N_z=16$.
}
\label{fig:spin-spin-sus}
\end{figure}

The numerical effort for a calculation of these retarded 
spin-spin susceptibilities and their spectral functions  using  the NRG
grows exponentially at larger distances  since the high-energy spectrum of the 
NRG response function is much less accurate than its low-frequency counterpart. While
the analytical calculation makes full use of the bath continuum, the conduction band 
is discretized on a logarithm energy scale \cite{Wilson75} and comprise only
a few bath sites representing the high energy spectrum.
Note that the NRG is geared towards the calculation of the impurity properties,
while we are using the NRG to extract a  bath correlation function.
Therefore, we are limited to short distances for  the bath spectral resolution 
in the NRG calculation.

To illustrate this point, we present a comparison between the analytical
and the NRG spectral function as a benchmark  in Fig.\ \ref{fig:spin-spin-sus}(a). 
While the agreement between the two results is very good at short distances, 
significant deviations are observed at $k_{\rm F}R/\pi=2.01$. 
Their origin can be traced to the limitation of the NRG to accurately 
resolve the high-energy part of the oscillations in the spectrum. 
The frequency scale of these oscillations is of the order of the band width $D$
as can be seen from Eq.\ \eqref{eq:spin-sus-fermi-sea}.
These high-energy oscillations cannot be properly resolved for large $R$
by the finite number of excitation energies
provided by the NRG in this frequency range for a finite $\Lambda>1$.  
The low-energy part of the spectrum, however is excellently recovered by the NRG
as expected.

After benchmarking the accuracy of the spectral functions at small distances, we have used 
the Fourier transformation Eq.\ \eqref{eq:sus-imp-bath} and the corresponding 
version for $\chi^r_{\rm c-c}(R,t)$ to calculate the retarded spin susceptibility in the time domain. 
A  time-dependent  and spatial conduction electron spin density $\expect{s^z(R,t)}$ 
is induced  as a response to a fictitious local 
Zeeman splitting $\Delta(t)=g\mu_{\rm B} B \theta(t)$ applied locally
at the origin. 

We compare $\expect{s^z(R,t)}$  obtained from 
the NRG retarded spin-spin susceptibility  
$\chi^r_{\rm c-c}(R,t)$  (dashed line) and the analytical susceptibility (solid line)
in Fig.\  \ref{fig:spin-spin-sus}(b) 
Both susceptibilities have been calculated 
using the Fourier transformation of the data depicted in Fig.\  \ref{fig:spin-spin-sus}(a) 
and substituting  the resulting $\chi^r_{\rm c-c}(R,t)$ for $\chi^r_{\rm imp-c}(R,t)$ in Eq.\ \eqref{eqn:27}.
We have plotted  $\expect{s^z(R,t)}$ normalized
to the Zeeman energy  $\Delta_0=g\mu_{\rm B} B$ to eliminate the trivial proportionality to the
applied field strength.

The induced time-dependent spin-density polarization  $\expect{s^z(R,t)}$ can be understood
as a response to a spin wave propagating with the speed $v_{\rm F}$ 
through the lattice and a consecutive fast equalization.  
The maximum of the spin wave is expected to be found at the time $\tau_l=R/v_{\rm F}$ 
indicated by the arrows in Fig.~\ref{fig:spin-spin-sus}(b).
The analytical response can be evaluated  at arbitrary distances $R$.
Indeed, the center of the propagating spin wave is located exactly at the time $\tau_l$ 
for larger distances $k_{\rm F}R/\pi\geq 2$---not explicitly shown here---while for shorter distances we 
observe a slight shift as depicted  
in Fig.~\ref{fig:spin-spin-sus}(b).
Some response is found  outside of the light cone related to the  finite width of the spin wave. 
This finite spatial resolution is directly linked to the finite momentum cutoff in the analytical formula defined by the restriction of the $k$-values to the 
first Brillouin zone: a sharp suppression of the signal outside the light cone would
require sending the momentum cutoff to infinity
as done in the analytical calculation of Ref.~\onlinecite{Medvedyeva2013}.

The spin-wave obtained from the NRG calculations is slightly faster originating 
from the shift of spectral weight to higher energies due to NRG 
spectral broadening \cite{BullaCostiPruschke2008} of the finite size NRG spectra 
as illustrated also in Fig.\ \ref{fig:spin-spin-sus}(a).

By comparing the numerical response with its 
analytical counterpart it is apparent that the
small oscillations around the long time limit of the spin-density polarization 
are not a numerical artefact due to the NRG discretization errors but related to the finite
bandwidth and the linear conduction band dispersion.

Furthermore, we find a significant deviation between the long-term limit
of the spinpolarization calculated analytically and the one using the convolution
of the NRG retarded susceptibility for $k_{\rm F} R/\pi = 0.01$. It is straightforward
to see that the stationary value  is determined by the integral over $\rho^r_{\rm c-c}(R,\w)/\w$. 
As discussed above
and depicted in Fig.\ \ref{fig:spin-spin-sus}(a), the finite resolution and the 
NRG broadening shifts some spectral weight to higher frequencies compared to the 
exact solutions, and, hence, we find a reduced value 
for $|\expect{s^z(R\to 0,\infty)}|$ since $\rho^r_{\rm c-c}(0,\w)/\w$ does not change sign. 
Once the spectral functions exhibit sign changes, broadening and finite size errors partially cancel
and the accuracy of $\expect{s^z(R,\infty)}$ increases for larger $R$.

\subsection{Retarded susceptibility $\chi^r_{\rm imp-c}(R,t)$ }

\begin{figure}[t]
\centering
  \flushleft{(a)}
  \includegraphics[width=0.5\textwidth]{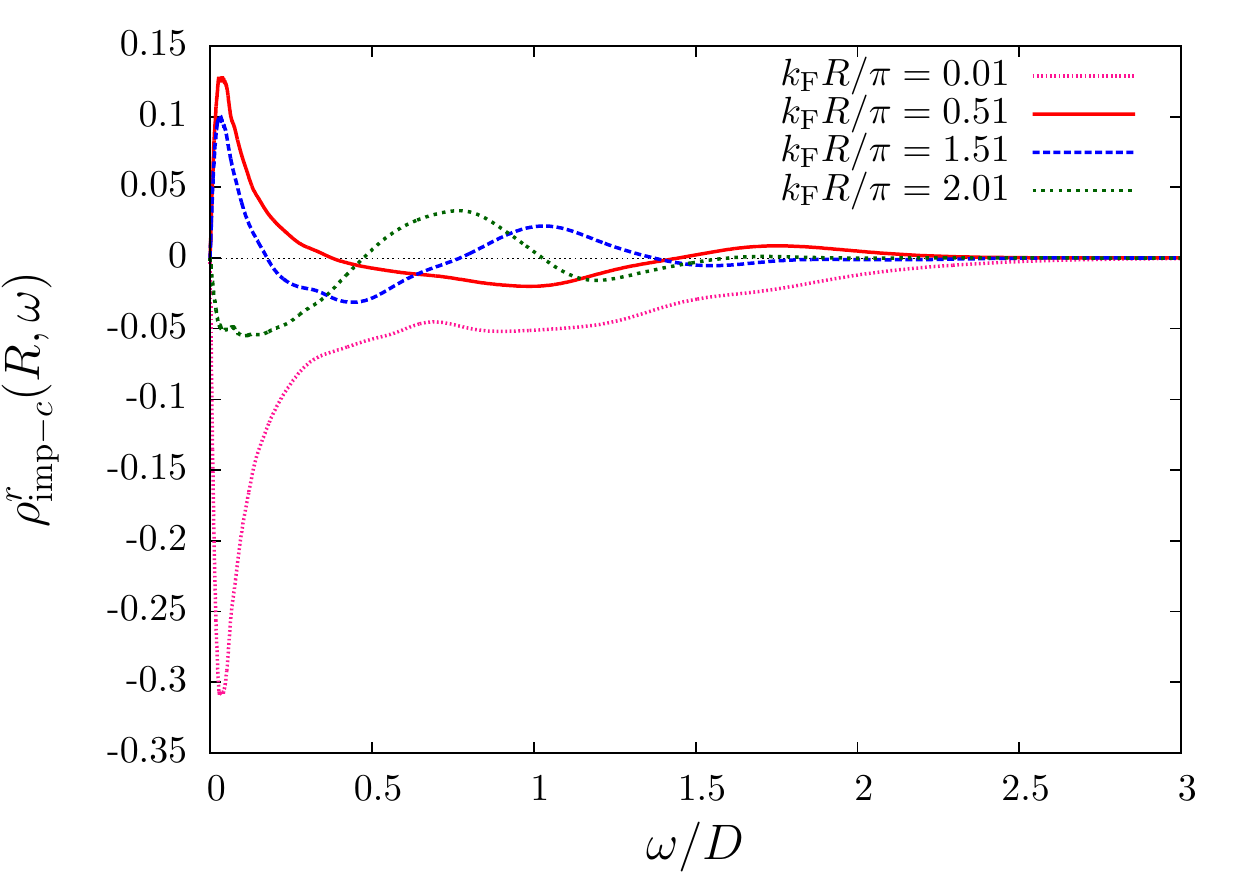}

  \flushleft{(b)}
  \includegraphics[width=0.5\textwidth]{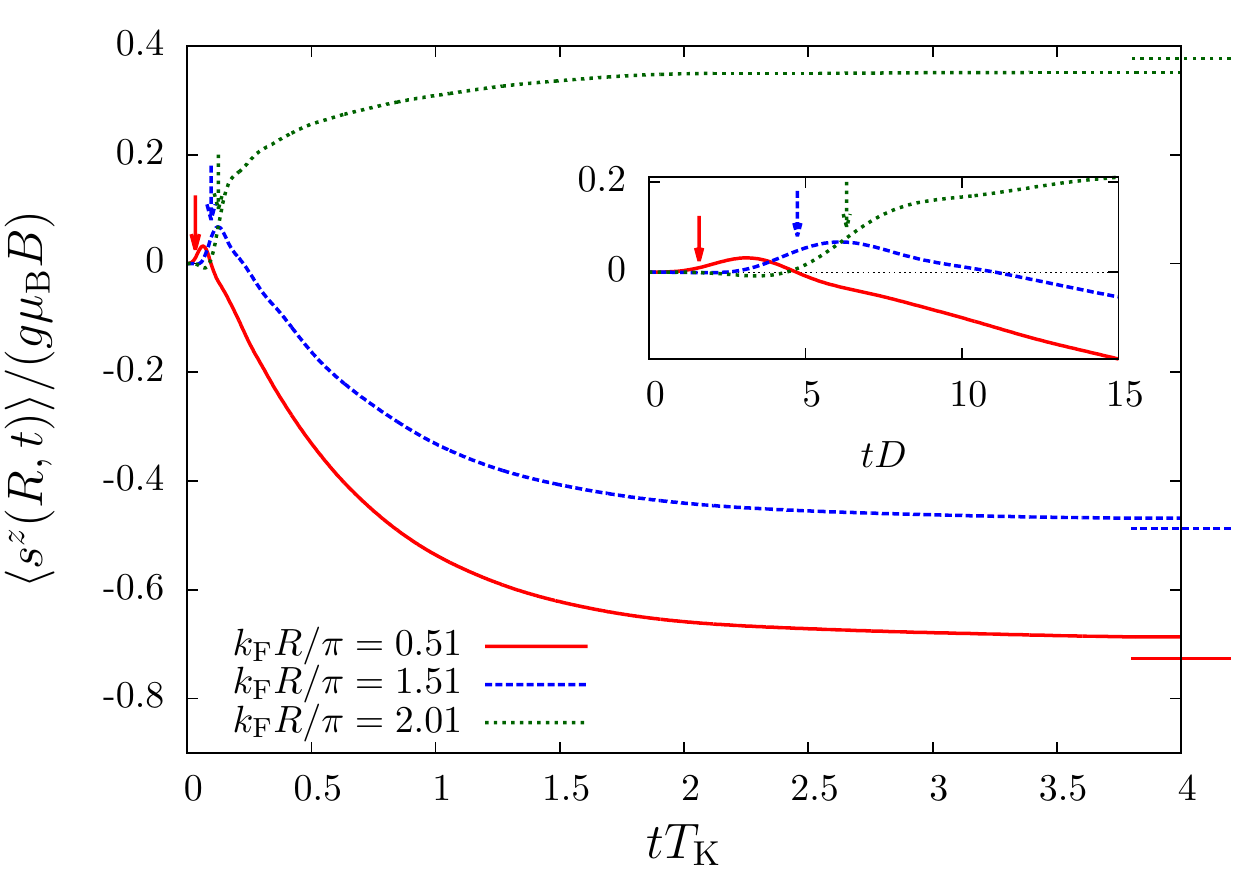}


\caption{(Color online)
(a) The  spectral function of  the retarded host-spin susceptibility 
$\rho^r_{\rm imp-c}(R,\w)$ for four different distances $R=0.01,0.51,1.51,2.01$.
(b) The time-dependent spin-polarization $\expect{s^z(R,t)}$ vs the dimensionless
time $t \, T_K$ after switching on the local magnetic Zeeman
splitting $\Delta(t)=g\mu_\mathrm{B}B\theta(t)$ on the impurity spin calculated using Eq.\ \eqref{eqn:27}
for three distances $0.51,1.51,2.01$.
The inset shows the short time behavior in more detail.
The horizontal lines on the right indicate the equilibrium
value $\expect{s^z(R,\infty)}/(g\mu_\mathrm{B}B)$
obtained by an equilibrium NRG
calculation with $N_z=1$ for a very small magnetic field  $g\mu_\mathrm{B}B/D=10^{-8}$
acting on the impurity spin.
NRG parameters are $\Lambda=2.25$, $N_s=3000$ and $N_z=16$ .
}
\label{fig:spin-spin-response}
\end{figure}

Now we proceed to the spin susceptibility $\chi^r_{\rm imp-c}(R,t)$ for finite $J$
describing the response of  the conduction band spin-density at some distance $R$ to
a perturbative magnetic field in $z$ direction coupling only to the impurity spin. 
Its spectral function is shown in Fig.\ \ref{fig:spin-spin-response}(a) 
for four different distances $R$ to illustrate the changes due to the presence of the Kondo spin ($J>0$).\\
Clearly, the increase of the numbers of oscillations in the frequency spectra 
with increasing $R$ prevails the  indication of the RKKY mediated spin response.
Furthermore, the antiferromagnetic coupling $J>0$ leads to a change of sign in $\rho^r_{\rm imp-c}$ compared to $\rho^r_{\rm c-c}$. 
A significant spectral weight is now located at low frequencies:
the  Kondo physics is reflected in  the distinctive peak  at $\omega \approx T_\mathrm{K}$ apparent for all distances $R$.

In Fig.\ \ref{fig:spin-spin-response}(b), the time dependence of the spin-density $\expect{s^z(R,t)}$ is shown 
for a Zeeman splitting $\Delta(t)=g\mu_\mathrm{B} B \theta(t)$ applied to the impurity spin.
Compared to  $\rho^r_{\rm c-c}$, one observes a change of sign in the response
due to the antiferromagnatic coupling.
For even multiples of $k_{\rm F} R/(2\pi)$, 
the conduction-band electron spin density aligns antiparallel, and for odd multiples the density
aligns 
parallel to the impurity spin in the long-time limit reflecting the RKKY mediated spin response.

We have identified two relevant time scales in the induced spin density  $\expect{s^z(R,t)}$: 
the fast light cone time scale $\tau= R/v_{\rm F}$ and the slow Kondo time scale $1/T_K$. 
The spin-density polarization remains almost zero until the 
ferromagnetic spin wave has propagated from the impurity spin to the distance $R$ after that 
$\expect{s^z(R,t)}$ starts building up [see inset 
of Fig.\ \ref{fig:spin-spin-response}(b)]. 
Again, the finite width of this  spin-wave response is directly linked to 
the finite momentum cutoff of our single symmetric conduction band used in the NRG
calculation as discussed above. The steady state, however, is reached very slowly: its sign is determined
by the $R$-dependent RKKY interaction, and  the long-time approach is governed by the Kondo scale $T_K$ independent of the distance 
as demonstrated in Fig.\ \ref{fig:spin-spin-response}(b).
This is in contrast to the fast response
of the decoupled Fermi sea, where the equilibrium spin polarization  
is reached rather fast  on the time scale $1/D$ as depicted in Fig. \ref{fig:spin-spin-sus}(b).

To gauge the quality of the long-time steady-state value obtained within the 
linear-response theory, we have used an equilibrium NRG approach
to calculate the equilibrium value $\expect{s^z(R,\infty)}/(g\mu_\mathrm{B}B)$ 
at a  very small magnetic field  $g\mu_\mathrm{B}B/D=10^{-8}$
acting on the impurity spin. These
values are added as a horizontal line on the right of Fig. \ref{fig:spin-spin-response}(b).
For vanishing local magnetic field, the linear response theory becomes exact,
and the steady-state value must coincide with the equilibrium expectation value in an infinitely 
large system.
Similar to  Fig \ref{fig:spin-spin-sus}(b), the steady-state value $|\expect{s^z(R,\infty)}|$ 
calculated from the spectral function is smaller than the equilibrium value caused by
the NRG broadening shift of spectral weight to higher energies. 
This origin of the deviations has already been discussed in Sec. \ref{sec:Retarded host susceptibility}.

\section{Discussion and outlook}
\label{sec:discussion-outlook}

We presented a comprehensive study of the spatial and temporal
propagation of Kondo-correlations for antiferromagnetic and
ferromagnetic Kondo couplings using the TD-NRG.

Our approach is based on a careful construction of two Wilson chains
obtained from the two distance dependent even and odd parity
bands. Full energy dependence and the correct normalization of the
bands are required for accurate results.  We have benchmarked our
mapping by (i) calculating the intrinsic spatial dependence of
spin-spin correlation of the Fermi sea, which coincides with the exact
analytical calculation for the full continuum, and (ii) checking the
sum rule of the equilibrium spin correlation function for
ferromagnetic and antiferromagnetic couplings. Our numerical data
fulfill the sum rule with an error of 1\% in 1D, which provided a
second independent check of the distance dependent NRG mapping.

The light cone defined by the Fermi velocity $R=v_{\rm F} t$ divides
the parameter space of the spatial and temporal correlation function
$\chi(R,t)$ in two segments. Inside the light cone, the
spin correlations develop rather rapidly, and then are followed by a much
slower decay towards the equilibrium
correlation function. Typical decaying Friedel oscillations with
a characteristic frequency $2k_{\rm F}$ are observed in the spatial
dependence. In the Kondo regime, the envelope function shows a
power-law behavior $1/R$ in equilibrium with ferromagnetic and
antiferromagnetic correlations for short distances and crosses over to
a $1/R^2$ behavior for distances exceeding the Kondo length scale
$R>\xi_{\rm K}$ in 1D. In this region only antiferromagnetic
correlations are found that correspond to a finite negative value of
the sum rule.

For the ferromagnetic regime, the inverse length scale $1/\xi_{\rm K}$
vanishes, and the correlation function remains oscillatory with a
slower decay of the envelope function.  The position of minima and
maxima are interchanged. The analytical calculation of the correlation
function provides an excellent understanding of our numerical TD-NRG
data.

Remarkably, we have found a building up of correlation even outside of the
light cone for ferromagnetic and antiferromagnetic Kondo couplings in
our TD-NRG data. We were able to trace back these correlations to the
intrinsic entanglement of the Fermi sea using a second-order
perturbation expansion in the Kondo coupling. The analytical structure
of the perturbative contribution provides an explanation of the
differences observed between ferromagnetic and antiferromagnetic Kondo
couplings.

The extension of the calculations to finite temperatures shows
that both correlations in- and outside of the light cone
are cut off at the thermal length scale $\xi_T=v_{\rm F}/T$.

Furthermore, we have presented the spectra of the retarded 
susceptibility for different distances $R$ and
have used  $\chi^r_{\rm imp-c}(R,t)$ to calculate the 
response of the spin-density polarization $\expect{s^z(R,t)}$
induced by a weak magnetic field that has been switched on locally at the origin
and interacting with the impurity spin.
For the real-time response of $\expect{s^z(R,t)}$, almost no correlations outside of the light cone were found
within the spatial resolution. The sharpness of the light cone is directly related to the momentum cutoff.

\begin{acknowledgments}
We would like to acknowledge fruitful discussions with Ian Affleck,
S.\ Kehrein and M.\  Medvedyeva.
This work was supported 
by the Deutsche Forschungsgemeinschaft under AN 275/7-1,
and supercomputer
time was granted by the NIC, FZ J\"ulich under project
No. HHB00.

\end{acknowledgments}

\appendix

\section{RKKY-INTERACTION}
\label{sec:appendix-RKKY}

In this appendix, we briefly summarize  
how the effective RKKY-interaction between two impurities mediated by a
single conduction band  is calculated in second order in $J$.

The interaction between the conduction band electrons 
and the impurities can be expanded in even $(e)$ and odd $(o)$ parity states \cite{Jones_et_al_1987,AffleckLudwigJones1995} and is given by
\begin{widetext}
\begin{eqnarray}
	H_\mathrm{int}&=&
	\frac{J}{8}\sum_{\sigma,\sigma'}\int \int d\epsilon d\epsilon' 
	 \sqrt{\rho(\epsilon)\rho(\epsilon')}\vec{\sigma}_{\sigma,\sigma'} \left[
	 \left(\vec{S}_1+\vec{S_2}\right) 
	 \right.
	 \left( N_e(\epsilon) N_e(\epsilon') c^\dagger_{\epsilon\sigma,e} c^{\phantom{\dagger}}_{\epsilon'\sigma',e} \right. 
	 \left. + N_o(\epsilon) N_o(\epsilon')  c^\dagger_{\epsilon\sigma,o} c^{\phantom{\dagger}}_{\epsilon'\sigma',o} \right) \nonumber \\ 
    && 
    \phantom{\frac{J}{8}\sum_{\sigma,\sigma'}\int \int d\epsilon d\epsilon' }
    + \left.
     \left(\vec{S}_1-\vec{S_2}\right) 
         \left(N_e(\epsilon)N_o(\epsilon') c^\dagger_{\epsilon\sigma,e} c^{\phantom{\dagger}}_{\epsilon'\sigma',o} + \text{H.c.} \right) 
    \right]
    \punkt \label{eq:H_int}
\end{eqnarray}
\end{widetext}

\begin{figure}[bt]
	\centering
	\includegraphics[width=0.25\textwidth]{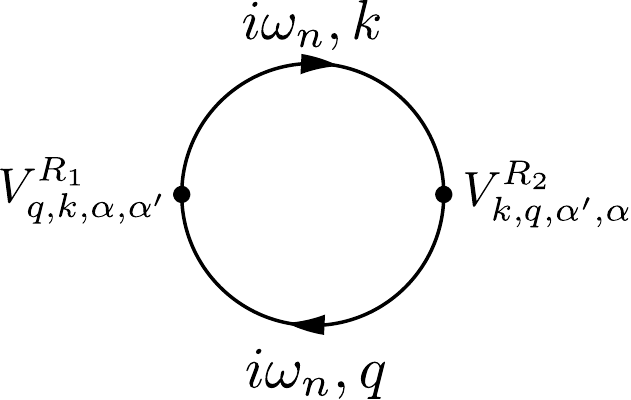}
	\caption{The second-order Feynman diagram generating the lowest-order 
	contribution to the RKKY interaction between to
	localized spins mediated by a spin excitation propagating 
	through the metallic host.
	}
	\label{fig:bubble}
\end{figure}

The normalization factors $\sqrt{\rho(\epsilon)} N_{e/o}(\epsilon)$ have been defined by
Eq.\  \eqref{eq:Norm_1D} in 1D, by \eqref{eq:Norm_2D} in 2D and by \eqref{eq:Norm_3D} in 3D.
The RKKY interaction is generated by a propagation of
spin excitation in the conduction band between the two impurities. The leading
second-order contribution to the RKKY interaction is depicted as a Feynman 
diagram in Fig.\ \ref{fig:bubble}.
Integrating out the conduction electrons leads to the effective RKKY interaction $H_\mathrm{RKKY}$:
\begin{align}
	H_\mathrm{RKKY}=&\frac{1}{\beta} \sum_{i\omega}\int\int d\e d\e' 
	\sum_{\alpha,\alpha'}G^0(\omega_n,\e) G^0(\omega_n,\e') \nonumber \\
  &\times V^{\vec{R}_1}_{\alpha,\alpha'}(\e,\e') V^{\vec{R}_2}_{\alpha`,\alpha} (\e',\e) \komma
\end{align}
which is exact in second order in $J$. The vertex operator $V^{\vec{R}_i}_{\alpha,\alpha'}(\epsilon,\epsilon')$
at the  position $\vec{R}_i$, of the impurity spin $\vec{S}_i$ originates from the Hamiltonian $H_\mathrm{int}$ in \eqref{eq:H_int} and is defined as 
\begin{eqnarray}
V^{\vec{R}_i}_{\alpha,\alpha'}(\epsilon,\epsilon')&=&c_{p,p'}^i\frac{J}{8} \sqrt{\rho(\epsilon)\rho(\epsilon')}N_p(\epsilon) N_{p'}(\epsilon') \vec{\sigma}_{\sigma,\sigma'} \vec{S}_i 
\nonumber \\
\end{eqnarray}
and 
depends on the combined spin-parity index 
\text{$\alpha=(\sigma,p)$} with spin $\sigma$ and parity $p$, and the sign factor
\begin{align}
c_{p,p'}^i=\begin{cases}
   -1  & \text{if } p \neq p'\text{ and } i=2\\ 
  1 & \text{otherwise}
\end{cases} \punkt
\end{align}
\text{$G^0(\omega_n,k)=[i\omega_n-\epsilon_k+i\delta]^{-1}$} denotes the Green's function of a free electron, 
$i\omega_n$ the fermionic Matsubara frequencies.
A textbook \cite{Mahan81} evaluation of the summation over the Matsubara frequencies yields
\begin{eqnarray}
\frac{1}{\beta} \sum_{i\omega}G^0(\omega_n,\e) G^0(\omega_n,\e')&=& 
\frac{f(\epsilon)-f(\e')}{\epsilon-\epsilon'},
\end{eqnarray}
where $f(\epsilon)$ labels the Fermi-Dirac distribution.
For $T=0$ and a particle-hole symmetric conduction band, we arrive at
\begin{align}
	H_\mathrm{RKKY}=&
	\sum_{\alpha,\alpha'}\int_{-D}^0 \; d\epsilon \int_0^D \; d\epsilon' \left( \frac{V^{R_1}_{\alpha,\alpha'}(\epsilon,\epsilon')V^{R_2}_{\epsilon',\alpha}(\epsilon',\epsilon)}{\epsilon-\epsilon'} \right. \nonumber \\
		&\phantom{\sum_{\alpha,\alpha'}\int_{-D}^0} 
		 \left. + \, \frac{V^{R_2}_{\alpha,\alpha'}(\epsilon,\epsilon')V^{R_1}_{\alpha',\alpha}(\epsilon',\epsilon)}{\epsilon-\epsilon'} \right)
		\punkt
\end{align}
After performing the spin and parity summations we obtain the effective
spin-spin interaction
\begin{align}
	H_\mathrm{RKKY}=&\int_{-D}^0 \; d\epsilon \int_0^D \; d\epsilon' \; \rho(\epsilon) \rho(\epsilon')\frac{J^2}{16} 
	  \nonumber \\
      &\left(  \frac{ N_e^2(\epsilon) N_e^2(\epsilon') + N_o^2(\epsilon) N_o^2(\epsilon')}{\epsilon - \epsilon'} \right. \nonumber \\
	& \left. - \frac{N_e^2(\epsilon) N_o^2(\epsilon') + N_o^2(\epsilon) N_e^2(\epsilon') }{\epsilon - \epsilon'}  \right) \; \vec{S}_1\vec{S}_2, 
\end{align}
which defines the effective RKKY interaction constant $K_\mathrm{RKKY}$ as
\begin{align}
	K_\mathrm{RKKY}=&\int_{-D}^0 \; d\epsilon \int_0^D \; d\epsilon' \; \rho(\epsilon) \rho(\epsilon') \frac{J^2}{16} \nonumber \\
& \left( \frac{ N_e^2(\epsilon) N_e^2(\epsilon') + N_o^2(\epsilon) N_o^2(\epsilon')}{\epsilon - \epsilon'} \right. \nonumber \\
	 &- \left. \frac{N_e^2(\epsilon) N_o^2(\epsilon') + N_o^2(\epsilon) N_e^2(\epsilon') }{\epsilon - \epsilon'} \right),
\end{align}
which implicitly depends on the distance via the energy-dependent parity densities  $N_e(\epsilon)$ and
$N_o(\epsilon)$.  If their energy dependence is replaced by a  constant 
\text{$\sqrt{\rho(\epsilon)}N_p(\epsilon)=\rho_0 N_p$}, the approximation 
of Jones and Varma \cite{Jones_et_al_1987}
\begin{align}
	\frac{K_\mathrm{RKKY}}{D}=-\frac{J^2\rho_0^2}{16}2\ln(2)(N_e^2-N_o^2)^2
\end{align}
is recovered.  This  $K_\mathrm{RKKY}$, however, remains ferromagnetic for all distances $R$ and is 
insufficient to account for the correct spatial dependent RKKY interaction. 
As pointed out by Affleck 
and coworkers \cite{AffleckLudwigJones1995},  maintaining the energy dependence is crucial for 
the alternating ferromagnetic and antiferromagnetic interaction between two impurity spins
as a function of increasing distance.

\section{PERTURBATIVE APPROACH OF SPIN-SPIN CORRELATION FUNCTION
$\chi(\vec{r},t)$}
\label{sec:perturbation-Simp-s-bad}

We divide  the Hamiltonian into two parts \text{$H=H_0+H_K$} with $H_0=\sum_{\sigma,\vec{k}}
\e_{\vec{k}}\, c^\dagger_{\vec{k}\sigma}c^{\phantom{\dagger}}_{\vec{k}\sigma}$, with the free conduction band 
dispersion $\e_{\vec{k}}$ and $H_K=J\Simp \vec{s}_c(0)$.

The time-dependent spin-correlation function
\text{$\chi(\vec{r},t)=\langle \Simp\vec{s}(\vec{r})\rangle(t)$}
can be calculated as
\begin{align}
\label{eq:B1}
	\langle \Simp\vec{s}(\vec{r})\rangle(t)=\textrm{Tr}\left[ \rho^I(t) \Simp\vec{s}^I(\vec{r},t)\right]
\end{align}
using the density operator $\rho^I(t)$
in the interaction picture, which is defined for any operator $A$ as
\begin{align}
	A^I(t)=e^{iH_0t}Ae^{-iH_0t}
	\punkt
\end{align}
Since the impurity spin  commutes with $H_0$, it remains time independent.
The real-time evolution of $\rho^I(t)$ can be derived from the von-Neumann equation
\begin{align}
	\partial_t \rho^I(t)=i\left[ \rho^I(t), H^I_K(t) \right],
\end{align}
which is integrated to
\begin{align}
\label{eq:B4}
	\rho^I(t)=&\rho_0+i\int_0^{t} \left[ \rho_0,H^I_K(t_1) \right] \; dt_1 \\
		  &- \int_0^{t}\int_0^{t_1} \left[ \left[\rho^I(t_2),H^I_K(t_2) \right],H^I_K(t_1) \right] \;dt_2\; dt_1 \nonumber
\end{align}
using the boundary condition $\rho^I(0)=\rho_0$. 
For an approximate solution in $O(J^2)$, we replace $\rho^I(t_2)$ by $\rho_0$ in the second integral.
Substituting \eqref{eq:B4} into \eqref{eq:B1} and cyclically rotating the operators under the trace
 yields
\begin{align}
	&\langle \Simp\vec{s}(\vec{r})\rangle(t)\approx \Tr{\rho_0\Simp\vec{s}_I(\vec{r},t)} \nonumber \\
					  &+ i \int_0^t \Tr{\rho_0\left[H^I_K(t_1),\Simp\vec{s}_I(\vec{r},t)\right]} dt_1  \\
					  &- \int_0^t \int_0^{t_1} \Tr{\rho_0\left[H^I_K(t_2),\left[H^I_K(t_1),\Simp\vec{s}^I(\vec{r},t)\right]\right]} dt_2  dt_1\nonumber
\end{align}
containing only expectation values that only involve the initial density operator $\rho_0$
in which the impurity spin and the conduction electrons factorize.
In the absence of a magnetic field the first term vanishes, and the
initial correlation function is zero at $t=0$.
The integral kernel of the first-order correction is given by
\begin{widetext}
\begin{align}
	&\Tr{\rho_0\left[H^I_K(t_1),\Simp\vec{s}^I(\vec{r},t)\right]}=  
   -\frac{3}{4}\frac{J}{V_uN^2}\sum_{\vec{k},\vec{q}}f(\epsilon_{\vec{k}+\vec{q}})\sin\left( \vec{q}\vec{r} +(\epsilon_{\vec{k}+\vec{q}}-\epsilon_{\vec{k}})(t_1-t) \right) 
   \label{FirstOrder} \punkt
\end{align}
\end{widetext}
For a linear dispersion in 1D, the argument of the sine contains 
 $(\epsilon_{k+q}-\epsilon_{k})=v_{\rm F} q$  contributions. The kernel remains
phase coherent on the light cone $q(r - v_{\rm F} t)$ and, therefore, generates the  response 
on this light cone line.

Calculating the commutator of the second order yields 
\begin{widetext}
\begin{align}
	&\Tr{\rho_0\left[H^I_K(t_2),\left[H^I_K(t_1),\Simp\vec{s}^I(\vec{r},t)\right]\right]} \nonumber \\ 
&= \frac{3}{8}\frac{J^2}{V_uN^3}\sum_{\vec{k},\vec{q}_1,\vec{q}_2}f(\epsilon_{\vec{k}+\vec{q}_1})f(-\epsilon_{\vec{k}-\vec{q}_2}) 
\left\{ \cos\left[\vec{q}_1 \vec{r}+(\epsilon_{\vec{k}+\vec{q}_1}-\epsilon_{\vec{k}-\vec{q}_2})t_1 \right. \right. 
     \left. \left. +(\epsilon_{\vec{k}-\vec{q}_2}-\epsilon_{\vec{k}})t_2 + (\epsilon_{\vec{k}}-\epsilon_{\vec{k}+\vec{q}_1})t\right] \right. \nonumber \\
&\phantom{=}+\cos\left[\vec{q}_2\vec{r}+(\epsilon_{\vec{k}+\vec{q}_1}-\epsilon_{\vec{k}-\vec{q}_2})t_1 \right.
    \left. +(\epsilon_{\vec{k}}-\epsilon_{\vec{k}+\vec{q}_1})t_2 + (\epsilon_{\vec{k}-\vec{q}_2}-\epsilon_{\vec{k}})t\right] \nonumber \\
&\phantom{=}-\cos\left[(\vec{q}_1+\vec{q}_2)\vec{r}-(\epsilon_{\vec{k}+\vec{q}_1}-\epsilon_{\vec{k}-\vec{q}_2})t \right.
    \left. -(\epsilon_{\vec{k}-\vec{q}_2}-\epsilon_{\vec{k}})t_2 -(\epsilon_{\vec{k}}-\epsilon_{\vec{k}+\vec{q}_1})t_1\right] \nonumber \\
&\phantom{=}\left.-\cos\left[(\vec{q}_1+\vec{q}_2)\vec{r}-(\epsilon_{\vec{k}+\vec{q}_1}-\epsilon_{\vec{k}-\vec{q}_2})t \right. \right.
    \left. \left. -(\epsilon_{\vec{k}}-\epsilon_{\vec{k}+\vec{q}_1})t_2 - (\epsilon_{\vec{k}-\vec{q}_2}-\epsilon_{\vec{k}})t_1\right] \right\}\punkt \label{SecOrder}
\end{align}
\end{widetext}
\twocolumngrid
Because of the simple sine and cosine structure, the time integration can be obtained
analytically.  For the momentum integrations over $\vec{k}$, $\vec{q}_1$ and $\vec{q}_2$ we insert 
a 1D linear dispersion for $\epsilon_{\vec{k}}$.
If we expand \eqref{FirstOrder} and \eqref{SecOrder} for small times around $t=0$,
the momentum integrations can also be calculated  analytically otherwise a numerical
integration has to be performed.

\section{RETARDED HOST SPIN-SPIN SUSCEPTIBILITY}
\label{sec:app-c}

In this section, we will analytically derive the  
spectral function $\rho^r_{\rm c-c}(R,\omega)$ of the retarded host spin-spin 
susceptibility 
\begin{align}
	\chi^r_{c-c}(R,t)=-i\expect{[s^z(R,t),s^z(0,0)]} \theta(t) \label{eqn:chi_c-c}
\end{align}
introduced in Eq.~\eqref{eq:spin-sus-fermi-sea}.
The spin-density operator $s^z(R,t)$, which is given by
\begin{align}
	s^z(R,t)=&\frac{1}{2V_uN}\sum_{k_1,k_2}\sum_{\alpha,\beta}\sigma^z_{\alpha,\beta}c^\dagger_{k_1,\alpha}c^{\phantom{\dagger}}_{k_2,\beta} \nonumber \\ 
		 &\times e^{-i(k_1-k_2)R}e^{i(\epsilon_{k_1}-\epsilon_{k_2})t}
\end{align}
in the Heisenberg picture, is inserted into the definition Eq. \eqref{eqn:chi_c-c}
yielding
\begin{align}
	\chi^r_{c-c}(R,t) =& \frac{-i\theta(t)}{4V^2_uN^2} 
	\sum_{\substack{k_1,k_2\\ k_3,k_4} }
	 \sum_{\substack{\alpha,\beta \\ \alpha',\beta'} }
	 \sigma^z_{\alpha,\beta}\sigma^z_{\alpha',\beta'}  e^{-i(k_3-k_4)R} \nonumber \\
			\times& e^{i(\epsilon_{k_3}-\epsilon_{k_4})t}
			\langle [c^\dagger_{k_3,\alpha'}c^{\phantom{\dagger}}_{k_4,\beta'},c^\dagger_{k_1,\alpha}c^{\phantom{\dagger}}_{k_2,\beta}] \rangle \punkt
\end{align}
The expectation value of the commutator can be simplified to
\begin{align}
	\langle [c^\dagger_{k_3,\alpha'}c^{\phantom{\dagger}}_{k_4,\beta'},c^\dagger_{k_1,\alpha}c^{\phantom{\dagger}}_{k_2,\beta}] \rangle= &
	\delta_{\alpha,\beta'}\delta_{\beta,\alpha'} \delta_{k_1,k_4} \delta_{k_2,k_3}\nonumber \\
	     &\times\left[f(\epsilon_{k_2})-f(\epsilon_{k_1})\right] \punkt 
\end{align}
After performing the spin and the $k_3,k_4$ summations, we arrive
at the closed form
\begin{align}
	\chi^r_{c-c}(R,t) =& \frac{-i\theta(t)}{2V_u^2N^2}\sum_{k_1,k_2} e^{-i(k_2-k_1)R}e^{i(\epsilon_{k_2}-\epsilon_{k_1})t} \label{eqn:response function}\nonumber \\
			&\times\left[f(\epsilon_{k_2})-f(\epsilon_{k_1})\right]
			 \punkt
\end{align}

Its Fourier transformation yields the analytic expression
\begin{align}
	\chi^r_{c-c}(R,z) =& 
	\frac{-i}{2V_u^2N^2}\sum_{k_1,k_2} \int_{0}^\infty e^{izt} e^{-i(k_2-k_1)R}e^{i(\epsilon_{k_2}-\epsilon_{k_1})t} \nonumber \\
			&\times \left[f(\epsilon_{k_2})-f(\epsilon_{k_1})\right] \;dt  \\
			=& \frac{1}{2V_u^2N^2}\sum_{k_1,k_2} 
			\frac{\left[f(\epsilon_{k_2})-f(\epsilon_{k_1})\right]e^{-i(k_2-k_1)R}}{z-(\epsilon_{k_1}-\epsilon_{k_2})},
\end{align}
where we have made explicit use of the $\theta$ function
and used a slightly imaginary frequency $z=\omega +i\delta$ and $\delta>0$  guaranteeing 
convergence of the Fourier integral.

Due to the complex phase factor $\exp[-i(k_2-k_1)R]$, the spectral function $\rho_{c-c}^r(R,\omega)$ defined as
\begin{eqnarray}
\rho_{c-c}^r(R,\omega) &=&
-\frac{1}{\pi} \lim_{\delta \rightarrow 0} \Im [\chi^r_{c-c}(R,\omega + i\delta)]
\end{eqnarray}
has two contributions: the first term is generated by a $\delta$ function stemming from
the $1/(z-\e)$ term and the second by $\Im  \exp[-i(k_2-k_1)R] = -\sin[(k_2-k_1)R]$.
To this end, we obtain
\begin{align}
	\rho_{c-c}^r(R,\omega) =& 
	\frac{1}{2\pi V_u^2N^2} \sum_{k_1,k_2} \left[f(\epsilon_{k_2})-f(\epsilon_{k_1})\right]
	  \nonumber \\
		        & \times \left[ \pi \cos[(k_2-k_1)R] \delta(\omega -(\epsilon_{k_1}-\epsilon_{k_2})) \right.\nonumber  \\
			& + \left. \frac{\sin[(k_2-k_1)R]}{\omega -(\epsilon_{k_1}-\epsilon_{k_2})} \right] \label{eqn:response spectrum} \punkt
\end{align}
Equation \eqref{eqn:response spectrum} contains the dimensionless frequency $k_F R$,
which is directly related to the  increasing oscillation with increasing distance $R$.


%

\end{document}